\documentclass[twocolumn,showpacs,preprintnumbers,amsmath,amssymb]{revtex4}

\usepackage{graphicx}
\usepackage{dcolumn}
\usepackage{bm}

\newcommand {\be}{\begin{equation}}
\newcommand {\ee}{\end{equation}}
\newcommand {\bea}{\begin{eqnarray}}
\newcommand {\eea}{\end{eqnarray}}
\newcommand {\bem}{\begin{displaymath}}
\newcommand {\eem}{\end{displaymath}}
\begin{document}

\preprint{ }

\title{ Phenomenology of Conduction in  Incoherent Layered Crystals }

\author{ George A. Levin \\}

\affiliation{Air Force Research Laboratory, Propulsion Directorate, 1950 Fifth Street, Bldg. 450, Wright-Patterson Air Force Base, OH 45433}



\date{\today}

\begin{abstract}
A novel phenomenological approach to the analysis of the conductivities of incoherent layered crystals is presented. It is based on the fundamental relationship between the resistive anisotropy $\sigma_{ab}/\sigma_c$ and the ratio of the phase coherence lengths in the respective directions. We explore the model-independent consequences of a general assumption that the out-of-plane phase coherence length of single electrons is a short fixed distance of the order of interlayer spacing. Several topics are discussed: application of the scaling theory, magnetoresistivity, the effects of substitutions and the intermediate regime of conduction when both coherence lengths change with temperature, but at different rate.  
\end{abstract}

\maketitle

\section{\label{sec:level1}Introduction\protect}

The normal state conductivity of highly anisotropic layered crystals  such as underdoped high-$T_c$ cuprates  exhibits a number of unusual features widely discussed in literature.   Metallic in-plane ($\sigma_{a}$) and non-metallic out-of-plane ($\sigma_c$) DC conductivities coexist in a broad ranges of temperature and doping\cite{Cooper}. A prominent feature of underdoped layered cuprates  is strongly temperature dependent  anisotropy  $\sigma_{a}/\sigma_c$, which in many cases shows no tendency to saturation even at low temperatures.  Optical conductivity $\sigma_c(\omega )$  is approximately frequency  independent  over a  wide range of frequencies. These and other findings have led to conclusion that the out-of-plane transport in these crystals is incoherent\cite{Anderson,H,Tsvetkov,Basov,Marel,Qiu}.

The microscopic models based on the assumption of incoherence of the c-axis transport such as that in Ref.\cite{Graf} and others\cite{Cooper,Anderson}  corroborate this conjecture, because the properties of $\sigma_c $ obtained within these models reproduce some of the features of the real systems.
Presently, however,  the microscopic models   are  still too restrictive and
idealized  to provide a comprehensive description of the phenomenon of
incoherent conduction and to be a  versatile  framework for the analysis
of the experimental data.

It would be highly useful to have a phenomenology of incoherent transport
that would rely minimally on the microscopic models and was based  mainly on the
exploration of the consequences of a  general assumption
that the {\it out-of-plane phase coherence length} of the charge carriers
{\it  is a short  fixed distance}
of the order of interatomic distances. A novel approach  presented here
is based in large part on the  relationship between resistive
anisotropy $\sigma_a/\sigma_c$ and phase coherence lengths of the charge carriers.

The scope of the paper can be summarized as follows: In Secs. II and III we have derived the relationship between the anisotropy and phase coherence lengths and compare the results with the solvable microscopic models. In Sec. IV the scaling theory is applied in order to develop a more comprehensive description of conductivities than that offered by existing microscopic models. 

When the out-of-plane phase coherence length $\ell_{\varphi,c}$ is fixed, the resistive anisotropy is T-dependent and reflects the T-dependence of the in-plane coherence length $\ell_{\varphi,a}$, namely $\eta\equiv (\sigma_a/\sigma_c)^{1/2}\propto \ell_{\varphi,a}$. Thus, the analysis of the dependence of conductivity on the anisotropy,  $\sigma$ vs $\eta$, as opposed to
conventional $-$ conductivity vs. temperature  $-$ approach, gives an opportunity to gain an insight into the dependence of the conductivities on the
coherence length. An extention of the one-parameter scaling hypothesis\cite{Abrahams,Thouless} on the conductance of the phase-coherent volume leads to an  experimentally verifiable prediction that there may exist a unifying description of conductivity of incoherent crystals  at different doping levels of the form $\sigma /\bar\sigma =f(\eta /\bar\eta)$, where $f(y)$ is a universal function for a given class of crystals, and $\bar\sigma$ and $\bar\eta$ are doping-dependent normalization constants. The scaling theory allows to obtain the functional form of this dependence for insulating and metallic branches of the trajectories. For the metallic branch $\sigma_{ab}\sim\ln\ell_{\varphi}$ and for the insulating branch $\rho_{ab}\sim\ln\ell_{\varphi}$. At low temperatures this translates into logarithmic temperature dependence of resistivity observed earlier\cite{Gant,Ono}. The logarithmic dependence of {\it conductivity} is also shown to be present in $Y_{1-x}Pr_xBa_2Cu_3O_x$ crystals.

In Sec. V we discuss magnetoresistivity. An important result is that in the temperature range where one can observe the magnetoresistance caused by  quantum interference\cite{Lee}, the relationship between anisotropy of incoherent crystals and the in-plane coherence length:  $\sigma_{ab}/\sigma_{c} \propto \ell_{\varphi,ab}^2$, can be verified experimentally. Since the crossover  field of magnetoresistance $B_{\varphi}$ also depends on the phase coherence length,
$B_{\varphi} \propto 1/\ell_{\varphi, ab}^2$, a combination of seemingly unrelated quantities - $B_{\varphi}\sigma_{ab}/\sigma_{c}$ - should remain constant, even though each of the factors, $B_{\varphi}$ and anisotropy, strongly change with temperature.

Section VI is devoted to the effects of elemental substitutions and disorder in incoherent crystals. The treatment addresses sometimes puzzling effect of substitutions for $Cu$ and radiation damage on resistivities. In Sec. VII we discuss the "semicoherent crystals" in which both coherence lengths are changing with temperature, but at different rates, so that the anisotropy is still $T-$dependent, but differently than in fully incoherent crystals where  $\ell_{\varphi,c}= const.$ Section VIII offers some suggestions on the type of future experiments that can shed light on the nature of confinement in cuprates.

\section{\label{sec:level1} The relationship between anisotropy and phase coherence length \protect}

Kubo-Greenwood formula relates the diagonal components of the conductivity tensor to the matrix elements of the respective components of the position operator\cite{Economu}:
\be
\sigma_{xx}(\omega ) =
\frac{e^2\pi}{\Omega }\sum_{\alpha}\sum_{\beta \neq \alpha}|x_{\alpha  \beta}|^2(f_{\beta}-f_{\alpha})\omega\delta(\epsilon_{\alpha}-\epsilon_{\beta}-\hbar\omega ).
\ee 
Here $\{x,y,z\}$ are the principal axes of the crystal, $\Omega$ is the volume of the system, $\omega$ is the frequency, and the static conductivity $\sigma_{xx}=\lim_{\omega\to 0}\sigma_{xx}(\omega ) $. $x_{\alpha \beta}$ are the matrix elements of the $x-$component of the position operator between states of energy $\epsilon_{\alpha},\epsilon_{\beta}$, whose probabilities of occupation in thermal equilibrium are given by $f_{\alpha},f_{\beta}$. Two other diagonal components of the conductivity tensor, $\sigma_{yy}$ and $\sigma_{zz}$, are determined, respectively, by the matrix elements $y_{\alpha\beta}$ and      $z_{\alpha\beta}$. 

The directional dependence of the conductivities is determined entirely by the matrix elements of the position operator. Therefore, it is obvious that one can always find the length scales $\{\lambda_x, \lambda_y, \lambda_z\}$ such that 
\begin{displaymath}
\sum_{\alpha\beta}|\langle\alpha |x/\lambda_x |\beta\rangle|^2=\sum_{\alpha\beta}|\langle\alpha |y/\lambda_y |\beta\rangle|^2 = \sum_{\alpha\beta}|\langle\alpha |z/\lambda_z |\beta\rangle|^2.
\end{displaymath}
The summation here is over the states relevant to conductivity $(1)$. Thus, the anisotropy is determined by the ratio of these length scales: 
\be
\frac{\sigma_{xx}}{\sigma_{yy}}=\frac{\lambda_{x}^2}{\lambda_{y}^2};\qquad
\frac{\sigma_{xx}}{\sigma_{zz}}=\frac{\lambda_{x}^2}{\lambda_{z}^2}.
\ee 
What is the physical nature of these characteristic lengths? If we look at the ways the Eq. (1) can be reduced to the standard result of kinetic theory \cite{Economu,Th}, the matrix elements in (1) are obtained by integrating over the phase-coherent volume ( which is, roughly speaking, a rectangular block with
the  sides equal to the phase coherence lengths in the respective directions). Therefore, at least in the case of Fermi liquid type system, the length scales determining anisotropy in Eq. (2) are the phase coherence lengths. 

A deeper insight into the meaning of the length scales in Eq.(2) can be gained from  Thouless' concept  of conductance of a microscopic block. The  conductance of a block  whose  edges are along  the principal axes of the conductivity tensor is given by \cite{Th,Abrahams,Thouless,Akkermans}:
\be
g_i= \frac{e^2}{\hbar}\frac{dN}{dE}\langle\Delta E\rangle_i,
\ee
where $i=\{x,y,z\}$, $dN/dE$ is the total number of states inside the block per unit  energy and $\langle\Delta E\rangle_i$ is
the mean fluctuation in energy levels caused by replacing periodic by antiperiodic boundary conditions  normal to the  direction of the current.

The sensitivity of the energy spectrum  to the boundary conditions depends on how well
the wave function  retains its phase coherence    along the path between the boundaries. We can choose the sides of the block such  that, on
average, the random phase acquired due to an inelastic interaction along the
way between the two opposite
boundaries  is the same, of the order of $\pi$, for all three pairs of the block boundaries.
Then, the effect of the  {\it imposed} phase difference between
the boundaries on the energy spectrum of such a block is isotropic; namely,
$\langle\Delta E\rangle_x  = \langle\Delta E\rangle_y = \langle\Delta E\rangle_z$.
According to Eq. (3)  the conductance of such a block is isotropic.
This  choice of the sides of the block
corresponds to the definition of the phase coherence length $\ell_{\varphi,i}$:
the distance over which electrons
lose phase coherence\cite{Lee}. Thus, the conductance  $g_{\varphi,i}$
of the phase-coherent volume of an anisotropic medium is isotropic:
\be 
g_{\varphi,x}=g_{\varphi,y}=g_{\varphi,z}\equiv g_{\varphi}.
\ee

The conductances of neighboring  phase-coherent volumes add up according to
Ohm's law \cite{Lee}. Therefore,  a
macroscopic block $\{L_x,L_y,L_z\}$  obtained by fitting together  $N^3$
phase-coherent volumes
$(L_x/\ell_{\varphi,x} =L_y/\ell_{\varphi,y}=L_z/\ell_{\varphi,z}=N\gg1)$
also has an isotropic conductance
$G\approx Ng_{\varphi}$, which   can be  expressed in terms of the
components of the conductivity tensor:
\be G=\frac{\sigma_{xx}L_yL_z}{L_x}=
\frac{\sigma_{yy}L_xL_z}{L_y}=\frac{\sigma_{zz}L_yL_x}{L_z}.
\ee
This leads to the following relationship between conductivities:
\be
\frac{\sigma_{xx}}{\sigma_{yy}}=\frac{\ell_{\varphi,x}^2}{\ell_{\varphi,y}^2};\qquad
\frac{\sigma_{xx}}{\sigma_{zz}}=\frac{\ell_{\varphi,x}^2}{\ell_{\varphi,z}^2}.
\ee
The Thouless' definition of conductance [Eq. (3)]  is  a very general
approach to dissipation, because it relates conductance only  to the statistical properties of the energy spectrum. Therefore, the {\it conjecture} expressed by Eqs. (4) and (6) may have a general applicability, regardless of the type of conduction. 

\subsection{\label{sec:level2} Comparison with Fermi liquid }

A characteristic property of  conventional Fermi liquids  is that the phase coherence of the wave function extends over many unit cells in all directions. The  coherence lengths in different directions change with temperature at the same rate, determined by the inelastic relaxation time which is equal, or scales with, the decoherence time $\tau_{\varphi}$.  As the result, the anisotropy of Fermi liquids is temperature independent.

Specificaly, in a Fermi liquid with ballistic transport   $\ell_{\varphi,i}\propto v_{F,i}\tau_{\varphi}$, so that  Eq. (6) reduces to:
\be
\frac{\sigma_{ii}}{\sigma_{jj}}=\frac{v_{F,i}^2}{v_{F,j}^2}.
\ee
Here $v_{F,i}$ is an average component of the Fermi velocity
and $\tau_{\varphi}$ is assumed to be independent of the  velocity. Equation (7) coincides with the well known result for Fermi liquids which  follows from the solution of the kinetic equation under the same assumption, that the decoherence time is independent of the velocity. In the case of the parabolic dependence of the energy upon momentum, Eq.(7) reduces to the ratio of the effective masses:
\be
\frac{\sigma_{ii}}{\sigma_{jj}}=\frac{m_j}{m_i}.
\ee
Thus, Eq.(6) is consistent with the Fermi liquid result for anisotropy. At first glance this statement appears counter-intuitive since the conductivity of Fermi liquid is proportional to the first power of inelastic relaxation time and, consequently, the first power of $\ell_{\varphi}$. But a closer look at the expression for anisotropic conductivity\cite{Abrikosov} immediately clarifies this misapprehension: $\sigma_{ii}\propto e^2\langle v_{F,i}^2\rangle\tau_{\varphi}$ can be written as $\sigma_{ii}\propto e^2 \ell_{\varphi,i}^2/\tau_{\varphi}$. Now all directional dependence of the conductivity is incorporated into $\ell_{\varphi,i}^2$. The relaxation time in the denominator cancels out in the anisotropy, which leads to Eq. (6). 

The comparison with Fermi liquids also allows  us to clarify the definition of coherence lengths in Eq. (6). It should be understood that $\ell_{\varphi,i}^2$ in Eq. (6) is the mean square average of the respective coherence length.

In case of diffusive motion ($\ell_{\varphi,i}^2\propto D_i\tau_{\varphi}$) Eq. (6) reduces to the Einstein relation $\sigma_i/\sigma_j=D_i/D_j$. The reverse procedure of obtaining Eq. (6) from the Einstein relation is not as straightforward and requires a microscopic model of conduction. See an example in Ref.\cite{Levin}. 

\section{\label{sec:level1} Incoherent crystals \protect}

Relationship (6) provides an effective tool for developing a phenomenological
description of electric transport in incoherent layered crystals. The nature of interlayer decoherence does not concern us here. It may be the result of the intralayer processes [when the rate of transitions between layers $\omega_{\perp}$ is small in  comparison  with the decoherence time ($\omega_{\perp}\tau_{\varphi}<1$)]
or interlayer transition itself (if a random phase of the order of $\pi$
is acquired during the transition from one plane to another). In either case,  the essential point is that the phase-coherent volume of such a crystal contains only one layer (bilayer in the case of cuprates like $YBa_2Cu_3O_x$).
Correspondingly, the out-of-plane phase coherence length,
hereafter denoted as  $\ell_0$,  is the $T-$independent distance
of  the order of interatomic (interlayer) spacing. Obviously, the coherence length cannot be arbitrarily small: it cannot be smaller than the size of the atomic orbitals. The immediate consequence of Eq. (6) is that in incoherent crystals

\be
\frac{\sigma_{ab}}{\sigma_c} = \frac{\ell_{\varphi}^2}{\ell_{0}^2}.
\ee
Here, for simplicity, we consider isotropic planes $\sigma_{xx}=\sigma_{yy}\equiv\sigma_{ab}$, and
$\ell_{\varphi,x}=\ell_{\varphi,y}\equiv \ell_{\varphi}$.

Since $\ell_{\varphi}$ monotonically increases with decreasing temperature,
so does the anisotropy. Thus, a fundamental feature of {\it incoherent crystals} is the  anisotropy which reflects the $T$-dependence of the in-plane phase coherence length.
A consequence of Eq. (9) is that   the out-of-plane  normal  state conductivity
of incoherent crystals is completely determined by the in-plane conductivity and  phase coherence length:
\be
\sigma_c =\frac{\sigma_{ab}\ell_{0}^2}{\ell_{\varphi}^2}.
\ee

An important theoretical result that confirms Eqs. (9) and (10) is the work of Graf, Rainer and Sauls \cite{Graf}. They  modeled a layered metal as a stack of 2D conducting planes coupled via incoherent interplane  scattering of charge carriers. The planes were treated as 2D Fermi liquid. In the limit when the decoherence time $\tau_{\varphi}$ is the same for in-plane and out-of-plane conductivities, the anisotropy obtained in\cite{Graf} is given by:
\be
\frac{\sigma_{ab}}{\sigma_{c}}=\frac{v_F^2\tau^2_{\varphi }}{4d^2},
\ee
where $d$ is the interplane distance and $v_F$ is the Fermi velocity of the circular Fermi surface. If we define $\ell_{\varphi}^2$ in Eq. (9) as the mean square average over the Fermi circle:
\begin{displaymath}
\ell_{\varphi}^2= v_F^2\tau^2_{\varphi}\int_{-\pi/2}^{\pi/2}\frac{d\phi}{2\pi} \cos^2\phi =\frac{v_F^2\tau^2_{\varphi}}{4},
\end{displaymath}
Eq. (11) reduces to Eq. (9) with $\ell_0=d$. If we take the planes to be  conventional two dimensional  Fermi liquids with 
$\sigma_{ab}= qv_F\tau_{in}\sim qv_F\tau_{\varphi}= q^{\prime}\ell_{\varphi}$, Eqs. (10) and (11) give $\sigma_{c}=q^{\prime}\ell_0^2/\ell_{\varphi}$, 
or $\sigma_c \sigma_{ab} =const=q^{\prime 2}\ell_0^2$. 

The comparison of Eq. (9) with Eq. (11) seems to indicate that when we apply Eq. (9) to the analysis of the experimental data in $YBa_2Cu_3O_x$, for example, we have to take  $\ell_0$ equal to the total distance between the bilayers ($12\AA$). However, the real crystals are not necessarily can be adequately described by a model of stacked metal planes. The analysis of magnetoresistivity given in Section {\bf VI} indicates that   $\ell_0$ in $YBa_2Cu_3O_x$ may be equal to the half of the distance between neighboring $CuO_2$ bilayers ($6\AA$). The decoherence, apparently, is taking place over the distance between a $CuO_2$ bilayer and the neighboring  $CuO$ chain layer.  The exact value of $\ell_0$ will become clear only when we understand the mechanism of decoherence in cuprates. More details on this matter are given in Section {\bf VII}. Hereafter I will take $\ell_0$ equal to the half distance between the neighboring $CuO_2$ bilayers. However, most of the results presented below do not depend on the exact value of $\ell_0$. 

Below we consider experimental data that corroborate the validity of Eqs. (9) and (10). It concerns two types of incoherent crystals: insulating and optimally doped cuprates.
\subsection{Hopping conduction in insulating layered crystals}

The first example which allows to test the validity of the conjecture (6) and its consequences (9) and (10) is the anisotropic hopping conduction. The coherence length of localized carriers is equal to the hopping distance. If the hopping distances in different directions are temperature independent, as is the case in the nearest neighbor hopping regime, or if they change with temperature at the same rate, which corresponds to anisotropic 3D variable range hopping, the anisotropy remains $T-$independent even though it may be very large. Exponentially strong $T-$ dependence of the conductivities cancels out.
It should be noted that the result equivalent to Eq. (6) was obtained earlier for the critical network model of the hopping transport\cite{Shklovskii}. It was shown that the anisotropy is given by the square of the ratio of the correlation lengths of the critical network. 

In incoherent layered insulating crystals like $PrBa_2Cu_3O_{7-\delta}$ the localized states are two-dimensional, i.e. comprised of the orbitals that all belong to one bilayer\cite{Levin}. In the variable range hopping (VRH) regime, the in-plane hopping distance increases with decreasing temperature, while in the out-of-plane direction the carriers advance in fixed steps equal to the distance between the neighboring bilayers. Thus, the resistive anisotropy of such crystals increases at low temperatures when VRH sets in. To determine the $T-$dependence of the anisotropy, let us consider the conventional picture of VRH where the average in-plane hopping distance can be found by maximizing the hopping probability
\be 
P(R)\propto \exp\left \{-2 \frac{R}{\lambda} -\frac{A}{{\cal N}(R^2-R_0^2)T}\right \}. 
\ee 
Here  $\lambda$ is the localization length, ${\cal N}=const$ is the 2D density of states, and $A$ a numerical coefficient. The only modification of the  traditional treatment of VRH is the denominator $R^2-R_0^2$ instead of $R^2$. This takes  into account that two localized states with close energies cannot be found closer to each other than a certain distance $R_0$ of the order of the localization length $\lambda$. Indeed, if the two states with close energies  strongly overlap, the phonon interaction that causes hopping will also hybridize them and push apart the energies of the new states. Equation (12) assures that in the limit of high temperatures the hopping distance does not decrease below the value $R_0$ which corresponds to the nearest-neighbor hopping. In the limit $T\rightarrow 0$, Eq. (12) crosses over into conventional Mott form. Strictly speaking, Eq. (12) is valid only for $R\gg R_0\sim 2\lambda$ which corresponds to VRH regime. In this limit the average hopping distance $\ell_h$, which in the hopping regime is equal to  the in-plane coherence length, is determined by the maximum of hopping probability $(12)$:
\begin{displaymath}
\ell_h\approx \bar R+ \frac{2R_0^2}{3\bar R};\qquad
\bar R = \left (\frac{\lambda A}{{\cal N}T}\right )^{1/3}
\end{displaymath}
Then, according to Eq. (9), 
\be
\frac{\sigma_{ab}}{\sigma_c}=\frac{\ell_h^2}{\ell_0^2}\approx \frac{4R_0^2}{3\ell_0^2} +
\left (\frac{\lambda A}{\ell_0^3{\cal N}T}\right )^{2/3}.
\ee
This T-dependence of the anisotropy  ($a+bT^{-2/3}$) was observed in insulating $PrBa_{2}Cu_{3}O_{7-\delta}$ and strongly underdoped superconducting $Y_{1-x}Pr_{x}Ba_{2}Cu_{3}O_{7-\delta}$\cite{Levin}. In  $PrBa_{2}Cu_{3}O_{7-\delta}$,  $a\approx 123$ and $b\approx 171\;K^{2/3}$. It is instructive to show that Eq. (13) not only determines correctly the temperature dependence of the anisotropy, but is quantitatively correct as well. The exponential temperature dependence of conductivity is determined by the maximum of the probability (12) ($\sigma\sim P_{max}$) at $R\approx \bar R$: 
\be
\sigma \sim \exp \left\{-\frac{3\bar R}{\lambda}\right\}=\exp \left\{-\left (\frac{T_M}{T}\right )^{1/3}\right \}; T_M=\frac{27A}{\lambda^2\cal N},
\ee
where $T_M$ is the Mott parameter. The anisotropy (13) can be presented as 
\be
\frac{\sigma_{ab}}{\sigma_c}=a + \left (\frac{T_a}{T}\right )^{2/3};\;\;\; T_a=\frac{\lambda A}{\ell_0^3\cal N}.
\ee
Thus,
\be
T_M=\frac{27 \ell_0^3}{\lambda^3}T_a
\ee

In Ref.\cite{Levin} we reported that $T_M\approx 300\;K$ and $T_a=b^{3/2}\approx 2236\;K$. This gives $\lambda = 3\ell_0 (T_a/T_M)^{1/3}\approx 6 \ell_0$. The constant term $a$ determines the cutoff length $R_0=\ell_0 (3a/4)^{1/2}\approx 9.6\ell_0=1.6\lambda$. This value of the cutoff $R_0$ is within  the expected range $\lambda <R_0<2\lambda$. For  $\ell_0\approx 6\AA$, we get $\lambda\approx 36\AA$. 
\subsection{\label{sec:level3} Resistivity of optimally doped high-$T_c$ cuprates}

As I already mentioned above, the assumptions that the in-plane conductivity of incoherent layered crystal is the Fermi liquid type: 
$\sigma_{ab}\approx e^2n\tau_{in}/m$ and the coherence length is proportional to the inelastic relaxation time $\tau_{in}$ lead to conclusion that the product $\sigma_c\sigma_{ab}$ must be temperature independent. Even though there is a great deal of evidence that optimally doped cuprates are incoherent, the temperature dependence of in- and out-of plane conductivities do not follow this rule. Since the anisotropy of such crystals directly determines the in-plane phase coherence length, Eq. (9), one can obtain the dependence conductivity vs. coherence length, $\sigma_{ab}(\ell_{\varphi})$, simply by plotting the conductivity data against anisotropy. In Ref.\cite{A1} the in-plane   conductivity of nearly optimally doped $YBa_2Cu_3O_{6.93}$ crystal was analyzed in this way over the temperature range  $ 90<T\le 300\;K$. The critical temperature of that sample was slightly below $90\;K$. The in-plane conductivity can be very accurately described as:
\begin{equation}
\sigma_{ab}=q( \ell_{\varphi}-\xi );\;\;\; \ell_{\varphi}>\xi,
\end{equation}
Here $\ell_{\varphi}\equiv \ell_0(\rho_c/\rho_{ab})^{1/2}$ and  $\xi\approx 21\AA$. This linear dependence $d\sigma_{ab}/d\ell_{\varphi}=const$  indicates ballistic motion of the quasiparticles. Indeed, the quasiclassical  conductivity  $\sigma_{ab}\propto \tau $
($\tau$ is the relaxation time of the distribution function). For ballistic motion $d\ell_{\varphi}/d\tau_{\varphi}=v_F$ ($\tau_{\varphi }$ is the decoherence time) and if $\tau \propto \tau_{\varphi }$,  then $d\sigma_{ab}/d\ell_{\varphi}\propto d\sigma_{ab}/d\tau_{\varphi
}=const$.According to  Eq. (10), the corresponding $\sigma_c$ is then given by
\begin{equation}
\sigma_{c}= q\ell_0^2 \left (\frac{1}{\ell_{\varphi}} - \frac{\xi }{\ell_{\varphi}^2}\right  ).
\end{equation}
Note that $d\sigma_{c}/d\ell_{\varphi} >0$ 
for $\ell_{\varphi} < 2\xi $, and $d\sigma_{c}/d\ell_{\varphi} <0$ for $\ell_{\varphi} > 2\xi $.  Since the phase coherence length monotonically increases with decreasing temperature,  $\sigma_{c}$ is metallic at high temperature ($\ell_{\varphi} < 2\xi$), reaches a maximum at a temperature $T$ where $\ell_{\varphi} (T) = 2\xi $, and
becomes nonmetallic, decreases, with further decreasing  $T$. The in-plane conductivity $\sigma_{ab}$ remains  metallic as long as Eq.(17) holds. 

Equations (17) and (18) can be rewritten as the following relationship between resistivities:
\begin{equation} (\rho_{c}\rho_{ab})^{1/2}= \bar\rho + \left (\frac{\xi}{\ell_0}\right )\rho_{ab};\;\; 
\bar\rho=\frac{1}{q\ell_0}.
\end{equation} 
A fit of the data with Eq. (19) gives  $\bar\rho \approx
0.23\;m\Omega\;cm $\cite{A1}.  Thus, the temperature dependence of
$\rho_{c}$ is determined by that of $\rho_{ab}$:
\begin{equation}
\rho_{c}=   \left (\frac{\xi}{\ell_0}\right )^2\rho_{ab}  +2\left (\frac{\xi}{\ell_0}\right )\bar\rho +
\frac{\bar\rho^2}{\rho_{ab}}.
\end{equation}
Note that Eq. (20) predicts a minimum in $\rho_{c}$ and an upturn at low temperatures. There is a characteristic value of $\rho_{ab}=\bar\rho \ell_0/\xi$, so that when $\rho_{ab}$ falls  below it, $\rho_c$ becomes nonmetallic. Since $\rho_{ab}$ is metallic, and decreases with decreasing temperature,  $\rho_c$ is metallic at high temperatures and nonmetallic at low temperatures. 
Specifically, when $\rho_{ab}=\alpha T$, which is characteristic of the optimally and nearly optimally doped cuprates,
\begin{equation}
\rho_{c}=  \beta_c + \alpha_c T +\frac{\gamma_c}{T}.
\end{equation}
The transition from metallic to nonmetallic $T$ dependence of $\rho_c$ with decreasing temperature is a characteristic feature of optimally doped and underdoped cuprates. In the crystals with maximum $T_c$ the onset of superconductivity sometimes can mask the upturn in $\rho_c$ so that it may look metallic at all temperatures above $T_c$\cite{Friedmann}. This is because, according to Eq. (18), $\rho_{c}$ becomes nonmetallic when $\ell_{\varphi} > 2\xi $ and it may not yet reach this value at temperatures above $T_c$. 

Using linear extrapolation of $\rho_{ab} = \beta +\alpha T$ with Eq.(20) one can predict where the  low temperature upturn in $\rho_c$, normally hidden by the onset of the superconductivity, begins. Experimentally, this minimum in $\rho_c$ can be revealed by the suppression of superconductivity by magnetic field similar to Refs.\cite{Ando,Ando1, Boebinger, Fournier,Ono}. A possible alternative to magnetic field application are the measurements of the normal state part of tunneling I-V curves in mesa junctions, see for example Refs.\cite{Krasnov,Yurgens}, provided that that the problem of Joule heating can be adequately addressed\cite{Yurgens1}. More detailed discussion of the application of intrinsic tunneling method to experimental exploration of the interlayer coherence of single electrons is given below in Sec. VII. 

In other types of crystals we may find the situation opposite to that in optimally doped $YBa_2Cu_3O_{7-y}$. In most experimental observations of the normal state resistivity the temperature range do not extend much above the room temperature and, in this range, $\rho_c$ may demonstrate the nonmetallic $T-$dependence. This is typical of the $Bi_2Sr_2CaCu_2O_{8+x}$ crystals. Optimally doped and even slightly overdoped crystals with $T_c$ in the range $80-90\;K$ have nonmetallic $\rho_c$ at all temperatures below room temperature. In contrast, the in-plane resistivity is metallic and increases with increasing temperature. According to Eq.(20) this situation corresponds to $\rho_{ab}<\bar\rho \ell_0/\xi$. However, this limit can be reached at sufficiently high temperature and beyond that $\rho_c$ will demonstrate metallic $T-$dependence. The pronounced  minimum in the $T-$dependence of $\rho_c$ in $Bi_2Sr_2CaCu_2O_{8+x}$ crystals have been reported in\cite{Yang} at $T\approx 750\;K$.  Thus, the apparently different temperature dependence of $\rho_c$ in optimally doped crystals such as  $Bi_2Sr_2CaCu_2O_{8+x}$ and  $YBa_2Cu_3O_{7-y}$ may, in fact, be the same, described approximately by Eq.(20), except that the minimum of $\rho_c$ is either hidden by the onset of superconductivity or lies above the temperature range within which the resistivity is measured.  

A nontrivial question raised by this analysis is the nature of the cutoff $\xi$  in the ballistic regime, Eq. (17). 
If we assume quasiclassical in-plane transport, namely,  $\sigma_{ab}\propto
\tau_{\varphi}$, the empirical Eq. (17)   gives $\ell_{\varphi}= q^{-1} \sigma_{ab} +\xi =v_F\tau_{\varphi} +\xi$. Therefore,  the cutoff   $\xi$  indicates  that the in-plane phase coherence length does not scale to zero with decreasing $\sigma_{ab}$ and  $\tau_{\varphi}$.  I emphasize that  this is the {\it high temperature} cutoff, important only when the decoherence time  is relatively short. On the other hand, if $\xi =0$ then  $\sigma_{ab}=q\ell_{\varphi}$, 
$\sigma_{c}=q\ell_0^2/\ell_{\varphi}$ 
and $\rho_c  \rho_{ab} =const= (q^2\ell_0^2)^{-1}$, instead of Eq.(19).
Without the finite cutoff $\xi$ in Eq.(17), the out-of-plane resistivity of incoherent crystals {\it cannot have metallic T-dependence} at high temperatures.

The data\cite{A1} that lead to Eq. (17) seem to indicate a phenomenon which can be  described as a "hard core" of the
phase-coherent volume. Even when $\sigma_{ab}\rightarrow 0$ at high temperature,  the in-plane phase coherence is retained within the area of the size
$\sim\xi^2\sim 20\times 20\AA^2$.
One might speculate  that such an unusually  large value of the coherence length at room temperature, accompanied by decreasing conductivity, 
is due to  backscattering resulted from static  disorder. In this sense, the "hard core"  of the phase-coherent volume of extended states may have  the same origin as  the  phenomenon of Anderson localization.

\section{\label{sec:level1} Scaling approach \protect}

A phenomenological description of conductivities that does not rely on specific microscopic models can be developed on the basis of   the one-parameter scaling
hypothesis\cite{Abrahams,Thouless}. It asserts that the  rate of change of the conductance of a microscopic block with its size depends only on the
conductance itself and nothing else.  By  extension,
the   derivative of the conductance of the phase-coherent volume
$d\ln g_{\varphi}/d\ln\ell_{\varphi}$
should also be a function of $g_{\varphi}$  only, namely:
\be
\frac{d\ln g_{\varphi}}{d\ln\ell_{\varphi}}=\kappa (g_{\varphi}/\bar g).
\ee
A nontrivial content of this statement  is that
the  other factors affecting conductance,
such as the concentration of dopants or impurities do not
change the functional dependence $\kappa (g_{\varphi}/\bar g)$, but may only affect  the normalization constant $\bar g$.
It was further suggested in\cite{Abrahams,Thouless}  that
$\bar g$ is universal ($\sim e^2/h$).
This assumption   seems to be too narrow and cannot hold
for all systems. A counter example is  a crystal with dopants that
change the number of carriers,
while the coherence length is dominated, let us say,
by inelastic electron-phonon  interaction.
Thus, the conductance changes with the concentration of dopants,
while $\ell_{\varphi}$
remains constant (at a given temperature).
For Eq. (22) to hold in this case, the normalization constant
must be concentration dependent.
If a crystal is incoherent, the general statement (22) about conductance
of the phase coherence
volume can be translated into experimentally  verifiable form.

According to Eq. (9), the anisotropy is a measure of the coherence length, $\ell_{\varphi} =\eta\ell_0$
[$\eta\equiv (\sigma_{ab}/\sigma_c )^{1/2}$].
On the other hand, the in-plane conductivity differs from
conductance $g_{\varphi}$ by a constant
factor, $\sigma_{ab}=g_{\varphi}/\ell_0$ (see Eq. (5)).
Thus, the observable consequence of the one parameter
scaling hypothesis is that
\be
\frac{d\ln \sigma_{ab}}{d\ln\eta}=\kappa (\sigma_{ab}/\bar \sigma).
\ee
In other words, the derivatives $d\ln \sigma_{ab}/d\ln\eta$   for a family of  crystals (such as  $YBa_2Cu_3O_{6+x}$, for  example) should be the same function of $\sigma_{ab}$ and should differ from each other only by the
$x-$dependent normalization constant $\bar \sigma$.
For the conductuctivity itself,  Eq. (23) translates into
a two parameter scaling:
\be
\frac{\sigma_{ab}}{\bar \sigma}= f\left (\frac{\eta}{\bar\eta}\right).
\ee
One can speculate that  $\bar \sigma $ changes  predominantly with changing   the density of carriers, while
$\bar\eta $ reflects mostly the degree of disorder present in the  crystal.
The scenario presented by Eq. (23) is illustrated in Fig. 1(a). The bold segments are  schematic representations of $\kappa(\sigma_{ab})$ for several  levels of doping. The range of $\sigma_{ab} $ corresponds to "accessible"  temperatures  $T_{min}<T<T_{max}$. ($T_{max}$ is typically $300-350\;K$ and $ T_{min}> T_c$ in superconducting crystals). The thin lines
indicate the hypothetical extensions of the trajectories
to "experimentally inaccessible" range of conductivities and temperatures.
\begin{figure}
\includegraphics{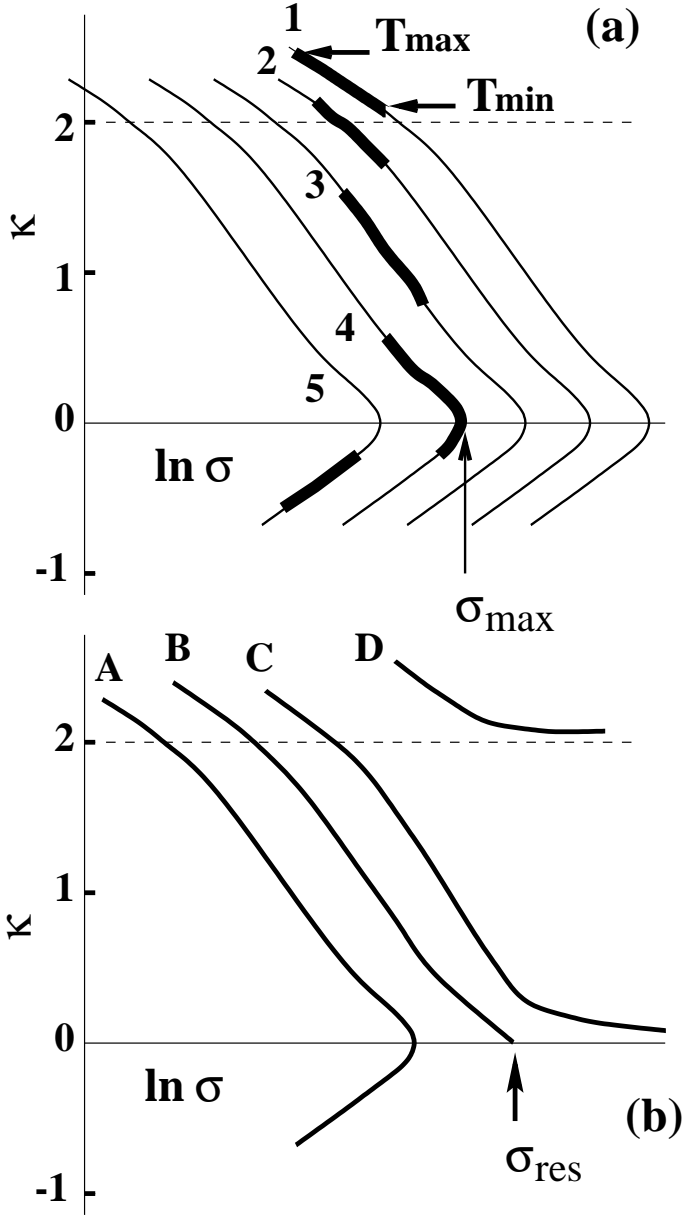}
\caption{\label{fig:epsart} {\bf (a)} Sketch of the trajectories $\kappa(\sigma_{ab} )$ vs.  $\ln\sigma_{ab}$. Bold segments  correspond to "experimentally accessible" values of conductivities as explained in the text. The thin lines are hypothetical extensions of the trajectories drawn under the assumption that the segments corresponding to different doping levels are parts of a continuous curve $\kappa(\sigma/\bar\sigma)$, shifted with respect to each other due to different normalization constants $\bar\sigma$. {\bf (b)} Sketch of the insulating ({\bf A}) and  metallic ({\bf C}) branches of the universal trajectories. Trajectory ({\bf B}) is the separatrix and ({\bf D}) is the hypothetical "supermetallic" branch.}
\end{figure}
One of the most interesting  points here is how the out-of-plane conductivity
evolves with doping. By definition,  
$\sigma_c(\eta ) =\sigma_{ab}(\eta )/\eta^2$
[see also Eq. (10)], so that
\bea
\frac{d\ln\sigma_c }{d\ln\eta }=\kappa -2.
\eea
Therefore, both $\sigma_{ab}$ and $\sigma_c$ are metallic for $\kappa>2$,
while metallic $\sigma_{ab}$ and  nonmetallic  $\sigma_c$ coexist for $0<\kappa<2$.
Both conductivities are   nonmetallic for $\kappa<0$.
Segment 1 in Fig. 1(a) represents a regime where both
conductivities $\sigma_{ab}$ and $\sigma_c$ are metallic
at all temperatures $T>T_{min}$,
because the whole segment is located above the threshold $\kappa =2$.

Segment 2 corresponds to a slightly underdoped system.
At high temperatures (smaller $\sigma_{ab}$) $\kappa >2$ and,  therefore,
$\sigma_c$ is metallic and increases with increasing $\eta\propto \ell_\varphi$; $\sigma_c$  reaches a maximum  when $\kappa (\sigma_{ab} )=2$ and decreases with further increasing $\ell_{\varphi}$. Therefore, at this level of doping $\sigma_c$ has a maximum ($\rho_c$  has a minimum) within the
accessible T-interval ($T_{min}<T<T_{max}$). The in-plane conductivity remains metallic.

Segment 3 represents a moderately underdoped system.
It is located  entirely within the range $0<\kappa<2$ and corresponds
to metallic $\sigma_{ab}$ and nonmetallic
$\sigma_c$ for all   $T_{min}<T<T_{max}$.

Segment 4 corresponds to strongly underdoped
crystals with the in-plane conductivity changing from
metallic at high T ($\kappa >0$) to nonmetallic at lower T ($\kappa <0$).
The singularity $\kappa =0$ is integrable:
\be
\kappa (\sigma )
\approx\pm \zeta\left (\ln\frac{\sigma_{max}}{\sigma}\right )^{1/2},
\ee
so that $\sigma_{ab}(\eta)$ determined by the equation
\be
\int_{\sigma_{max}}^{\sigma_{ab}}\frac{d\ln\sigma}{\kappa (\sigma
)}=\ln\frac{\eta}{\eta_1}
\ee
reaches the maximum value $\sigma_{max}$ at a finite $\eta=\eta_1$ and
decreases with further increasing $\eta$ (decreasing temperature).
Equations (26) and (27) give
$\sigma_{ab}(\eta)\approx \sigma_{max}\exp\{-\zeta^2\ln^2(\eta/\eta_1 )/4\}$.

Finally, segment 5 corresponds to an insulating crystal with nonmetalllic
$\sigma_{ab}$ and $\sigma_{c}$.
As shown, all five  curves in Fig. 1(a) represent the same dependence
$\kappa (\sigma/\bar\sigma )$ shifted with respect to each other, when plotted against $\log\sigma$, because of the different values of $\bar\sigma $ which is
determined by the  density of carriers $n$ and the level of disorder.
The reduction of $n$ by doping reduces $\bar\sigma $, shifting the respective
segment of $\kappa (\sigma/\bar\sigma )$ to lower  values of
$\sigma$ and, at the same time, ahead along the trajectory in terms of the
reduced variable $\sigma/\bar\sigma$.
Thus, the experimental data plotted as $\sigma$ vs anisotropy would
present the "snapshots"   of different stages of  evolution of a given
system along  the common trajectory.
For example, {\it close to optimal doping},
$\rho_c$ is  metallic in $YBa_2Cu_3O_{x}$\cite{Friedmann}, presumably because the minimum in
$\rho_c$ is hidden by the onset of superconductivity.  The decrease of oxygen content shifts the
"observable part" of the trajectory  so that it crosses the threshold $\kappa=2$ and the minimum in $\rho_c$ becomes evident at $T_c<T<300\;K$. This corresponds to $YBa_{2}Cu_{3}O_{x}$  with $x\approx 6.88$ and $6.78$ which   exhibit a minimum in $\rho_c$ at $T\approx 150$ and $300\;K$, respectively\cite{Takenaka}.
At even lower oxygen content, the minimum in $\rho_c$ shifts above the room temperature and
$\rho_c$ {\it appears} nonmetallic at all temperatures below $300\;K$.

A single universal trajectory shown in Fig. 1(a) predicts that, as temperature decreases, the incoherent crystals at all doping levels would eventually become insulating. First, $\rho_c$ becomes nonmetallic when $\kappa < 2$ and then, at even lower temperature, $\rho_{ab}$ turns nonmetallic ($\kappa <0$). Support for this scenario can be found in the experiments where superconductivity was suppressed by the magnetic field to reveal the underlying normal state. In the samples of $Bi_2Sr_2CuO_y$\cite{Ando1} and 
$Bi_2Sr_{2-x}La_xCuO_{6+\delta}$\cite{Ono} the prominent upturn in $\rho_{ab}$ has been revealed below $T_c$. The same phenomenon takes place in $La_{2-x}Sr_xCuO_4$\cite{Boebinger} and $Pr_{2-x}Ce_xCuO_4$\cite{Fournier}.

The results of the search for the universal trajectory described by Eq.(24) was reported in Ref.\cite{Euro}.  We have found that when the conductivity is plotted vs. anisotropy, the appropriately normalized data for several crystals of $ YBa_2Cu_3O_{6+x}$ and $ Y_{1-x}Pr_xBa_2Cu_3O_{7-\delta}$ with widely different composition fall on the same curve.

The analysis of the experimental data also suggest that one type of  trajectory $\kappa (\sigma )$ shown in Fig. 1(a) cannot describe all incoherent crystals.
Just like the genuinely 2D system\cite{Kravchenko}, the incoherent crystals may undergo metal- insulator transition. This scenario is schematically described in Fig. 1(b) by the trajectories {\bf A, B} and {\bf C}. The trajectory {\bf A} is the same as in Fig. 1(a) and corresponds to the insulating phase. The trajectory {\bf C} describes the metallic branch, and {\bf B} is the separatrix. The notion of insulating and metallic phase refers only to the in-plane resistivity. The out-of-plane resistivity is nonmetallic in either case as long as $\kappa < 2$. 
\begin{figure*}
\includegraphics{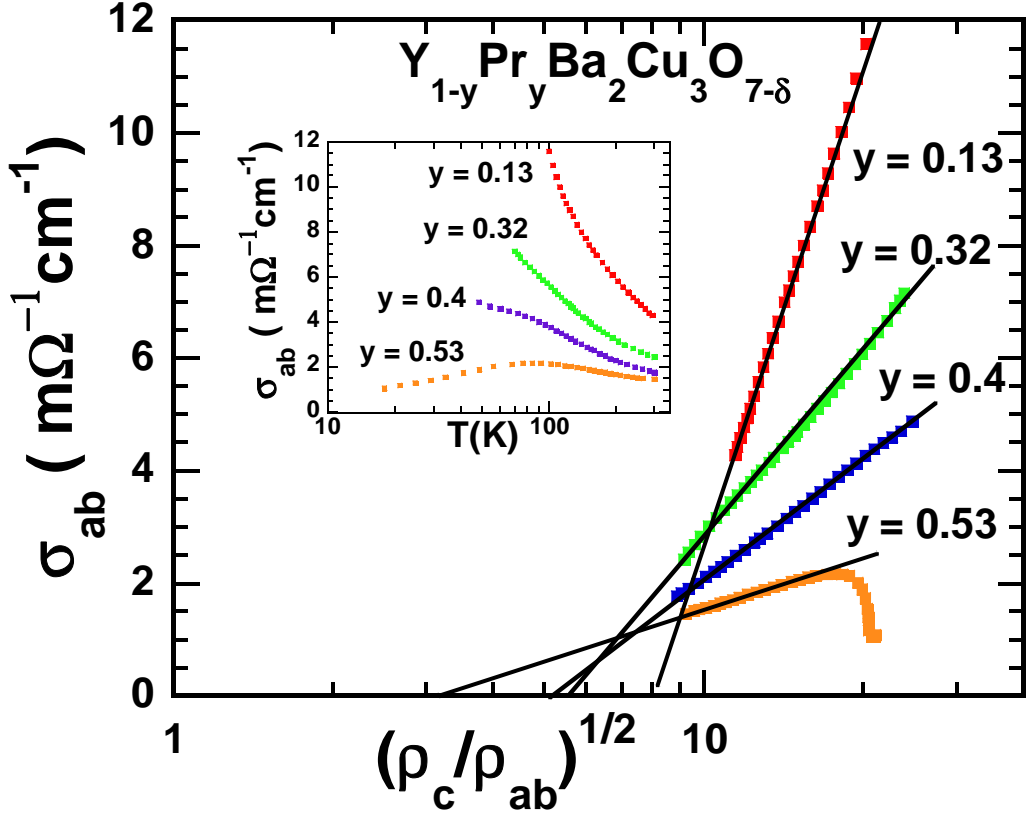}
\caption{\label{fig:wide} Conductivity of $Y_{1-y}Pr_yBa_2Cu_3O_{7-\delta}$ single crystals vs. logarithm of anisotropy. The solid lines are guide to the eye that correspond to $\sigma_{ab}\propto \ln\eta$ dependence. The inset shows conductivity vs. logarithm of temperature.}
\end{figure*}
In the limit $\kappa\to 0$ the separatrix can be approximated as 
$\kappa \approx \alpha\ln\sigma_{res}/\sigma$. Then, Eq. (27) gives
\be
\sigma_{ab}=\sigma_{res}\exp\{-\eta^{-\alpha}\}\approx \sigma_{res}(1- 
\eta^{-\alpha})
\ee 
Taking into account the relationship between the anisotropy and coherence length, one can rewrite the Eq. (28) as follows:
\be
\rho_{ab}=\rho_{res}\exp\left\{\left (\frac{\ell_0}{\ell_{\varphi}}\right )^ {\alpha}\right\}
\approx \rho_{res}\left (1 + \left (\frac{\ell_0}{\ell_{\varphi}}\right )^ {\alpha} \right )
\ee  
The out-of-plane resistivity diverges in the same temperature range. From Eq. (28) folows:
\be
\rho_{ab}=\rho_{res}\exp\left\{\left (\frac{\rho_{ab}}{\rho_{c}}\right )^ {\alpha /2}\right\},
\ee
which gives
\be
\rho_c=\frac{\rho_{ab}}{[\ln (\rho_{ab}/\rho_{res})]^{2/\alpha}}\approx
\frac{\rho_{res}^{1+2/\alpha }}{(\rho_{ab}-\rho_{res})^{2/\alpha}}.
\ee

Typically, the anisotropy $\eta$ is very large and $\rho_{ab}$ described by Eq. (28) is practically temperature independent, equal to the residual value        $\rho_{res}=\sigma_{res}^{-1}$. One of the samples of $Bi_2Sr_2CuO_y$ studied in Ref.\cite{Ando1} appears to exhibit the properties of the "separatrix crystal". Its in-plane resistivity remains  temperature independent between $11\;K$ and $0.7\;K$. The out-of-plane resistivity in the same range of temperatures diverges.    

Next, let us consider the metallic branch, the trajectory {\bf C} in Fig. 1(b). 
It describes $\rho_{ab}$ that is metallic at all temperatures ($\kappa >0$ for all values of $\sigma_{ab}$).
In incoherent crystals the phase-coherent volume is "two-dimensional"
(contains only one bilayer) and therefore, in the limit
$\ell_{\varphi}\rightarrow \infty$, the conductance should become size independent and, correspondingly, $\kappa\rightarrow 0$ . Let us assume that
$\kappa$ can be approximated as
\be
\kappa(\sigma)\approx \bar\sigma/\sigma.
\ee
In a different context a similar asymptotic dependence of the logarithmic derivative was discussed in Ref.\cite{Dobro}.
Integration of Eq. (27) leads to 
\be
\sigma_{ab}=\bar\sigma\ln (\eta/\eta_0), 
\ee
where $\eta_0$ is a constant of integration ($\eta>\eta_0$).
Given the definition of $\eta= (\rho_c/\rho_{ab})^{1/2}$, this translates into
the following relationship between resistivities:
\be
\rho_c=\eta_0^2\rho_{ab}\exp\left \{\frac{\bar\rho}{\rho_{ab}}\right \};\;\;
\bar\rho= 2\bar\sigma^{-1}.
\ee
Both parameters $\eta_0$ and  $\bar\rho$ are doping dependent. While $\rho_{ab}$ is metallic at all temperatures,  $\rho_c$ reaches  minimum at $\rho_{ab}=\bar\rho$. Depending on where $\bar\rho$ lies with respect to the range of measured $\rho_{ab}$,
we will see  all three types of $\rho_c(T)$ dependence  discussed above.
At sufficiently low temperatures the temperature dependence of the coherence length is a power law, which translates into the logarithmic increase of the conductivity:
\be
\sigma_{ab}=\bar\sigma\ln (\ell_{\varphi}/\ell_0\eta_0 )\sim
\ln (T_0/T), 
\ee
while the out-of-plane resistivity $\rho_c$ {\it diverges} as the power law.

\begin{figure*}
\includegraphics{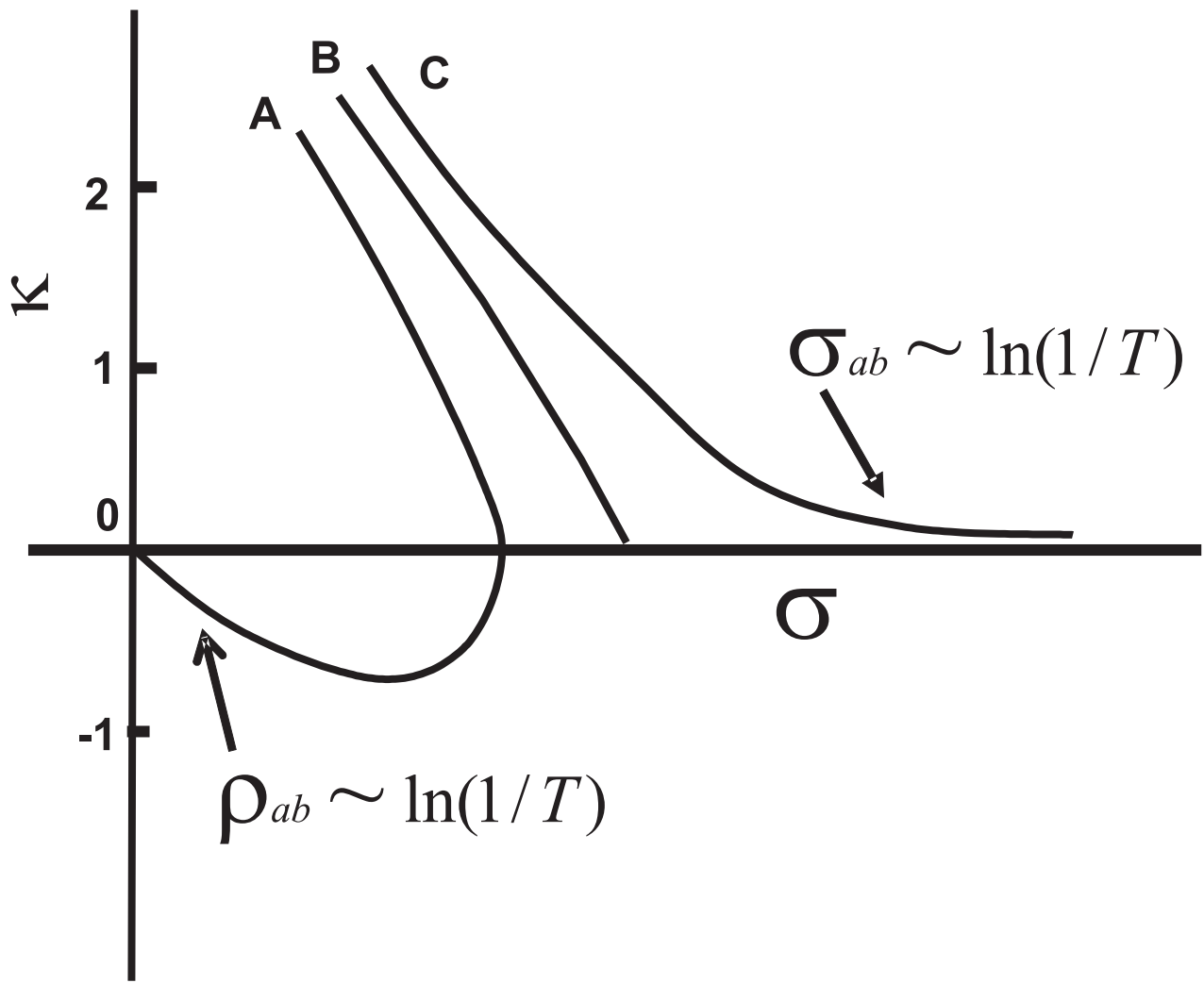}
\caption{\label{fig:wide} Sketch of three branches of the scaling trajectories: insulating ({\it non-hopping}) ({\bf A}), metallic ({\bf C}) and separatrix ({\bf B}). Unlike Fig.1, $\kappa(\sigma)$is shown vs. $\sigma$, not $\ln\sigma$.}
\end{figure*}
In Fig. 2 the logarithmic dependence given by Eq. (33) is illustrated by the data obtained on several samples of $Y_{1-y}Pr_yBa_2Cu_3O_{7-\delta}$.  This is the same data as in Ref.\cite{Euro}. In \cite{Euro} it was presented as $\ln\sigma$ vs. $\ln\eta$ in order to demonstrate the scaling given by Eq. (24), but such presentation  obscured the functional dependence $f(\eta)\propto \ln\eta $. The inset to Fig. 2 shows the conductivity of the same crystals vs. logarithm of  temperature. It is obvious that the conductivity cannot be described by the logarithmic temperature dependence. This is because the $T-$dependence of $\ell_{\varphi}$ in this temperature range has not yet settled into power law, but the trajectory $\kappa(\sigma)$ is already well described by the asymptotic dependence (32). Apparently, even the $\kappa >0$ part of the trajectory of the insulating sample $Y_{0.47}Pr_{0.53}Ba_2Cu_3O_{7-\delta}$ is well described by Eq. (32).  

Let us now turn to the insulating branch, the trajectory {\bf A} in Fig. 1(b). Usually it is assumed that in the insulating regime the mechanism of conduction is hopping: $\sigma\sim\exp\{-\chi\ell_{\varphi}\}$, which leads to 
$\kappa (\sigma)=\ln (\sigma/\sigma_0)$. There is, however, another possibility. In 2D insulator the limiting value of $\kappa(\sigma)$ may be zero, just as for the metallic branch: $\kappa(\sigma)\to 0$ as $\ell_{\varphi}\to\infty$ and $\sigma\to 0$. In Fig. 3 the alternative trajectory of insulating (non-hopping) branch ({\bf A}) that illustrates this scenario is shown. 
Assuming that $\kappa(\sigma)$ is an analytical function, the simplest option is
\be
\kappa(\sigma )\approx -\frac{\sigma }{\sigma_0}
\ee
when $\sigma\to 0$.
 This leads to 
\be
\rho_{ab}= \rho_0\ln (\eta/\eta_0).
\ee
At low temperatures, when $\ell_{\varphi}\sim T^{-\gamma}$, resistivity will acquire the logarithmic $T-$dependence:
\be
\rho_{ab}=\rho_0 \ln (\ell_{\varphi}/\ell_0\eta_0 ) \sim \ln (T_0/T).
\ee
This logarithmic $T-$dependence of the resistivity observed in $Bi_2Sr_{2-x}CuO_{6+\delta}$\cite{Ono} and $YBa_2Cu_3O_{7-y}$\cite{Gant} has attracted a great deal of attention and has not been explained by any microscopic model. 
The trajectory {\bf A} in Fig. 3 describes the in-plane resistivity that has metallic temperature dependence at high temperature, reaches minimum when $\kappa =0$ and logarithmically diverges at low temperature. The out-of-plane resistivity diverges as a power of temperature\cite{Gant}. The trajectories {\bf B} and {\bf C} in Fig. 3 describe the separatrix  and metallic branch, respectively, that are the same as in Fig. 1(b).

The last trajectory ({\bf D}) shown in Fig.1(b) describes a "supermetallic"  branch, such that both $\rho_c$ and  $\rho_{ab}$  are metallic at all temperatures ($\kappa >2$). If we take the asymptotic behavior of $\kappa\approx 2+\sigma_0/\sigma$, the resulting 
\be
\sigma_{ab}= \frac{\sigma_1}{2}\left (\frac{\eta}{\eta_0}\right )^2-\frac{\sigma_0}{2}.
\ee
Here $\sigma_1$ and $\eta_0$ are constants of integration. Using the definition of $\eta$ we get
\be
\rho_c=\rho_{min}+\gamma \rho_{ab}.
\ee
While $\rho_{ab}\to 0$, $\rho_c$ tends to a finite value. The asymptotic value $\kappa =2$ corresponds to diffusive in-plane motion of carriers. If we take the quasiclassical $\sigma_{ab}\sim\tau_{\varphi}$ and $\ell_{\varphi}^2\sim D\tau_{\varphi}$, then  $\sigma_{ab}\sim \ell_{\varphi}^2 $, which corresponds to $\kappa=2$. 

\section{Magnetoresistivity}
The effect of the magnetic field on resistivity of the incoherent crystals can be straightforwardly deduced from Eqs.(9) and (10) as long as we know what effect the magnetic field has on the coherence length. Hereafter I will discuss only the orbital part of the magnetoresistivity. Spin-dependent and orbital contributions can be separated due to their different field dependences\cite{Harris,Cimp,Cimp1}.  Normally, the application of the magnetic field perpendicular to the conducting layers decreases $\ell_{\varphi}$\cite{Lee}. The first obvious observation that follows from Eq.(9) is that {\it magnetoanisotropy} is always negative: 
\be
\frac{\Delta(\rho_c/\rho_{ab})}{\rho_c/\rho_{ab}}=
\frac{2\Delta\ell_{\varphi}}{\ell_{\varphi}}<0.
\ee
Magnetoanisotropy can be measured directly and, in fact, more accurately than the separate magnetoresistivities $\Delta\rho_{ab}$ and $\Delta\rho_{c}$ by the six-point method\cite{Cimp,JAP}. Even when both coherence lengths (in-plane and out-of-plane) change with temperature, the Eq.(41) holds as long as the the coherence length in the direction of the applied field does not change by the field, see Eq.(6). Indeed, in Refs.\cite{Ando1,Cimp} the negative magnetoanisotropy was observed in all samples of $Bi_2Sr_2CuO_y$ and $YBa_2Cu_3O_{7-\delta}$. 

Since both conductivities, $\sigma_{ab}$ and $\sigma_{c}$  are determined by the in-plane coherence length, Eqs.(23) and (25), there is a direct relationship between the type of conduction (metallic or nonmetallic) and the sign and magnitude of the magnetoresistivities (MR):
\be
\frac{\Delta\rho_{ab}}{\rho_{ab}}= -\kappa \frac{\Delta\ell_{\varphi}}{\ell_{\varphi}};\;\;\;\;\frac{\Delta\rho_{c}}{\rho_{c}}= -(\kappa -2)\frac{\Delta\ell_{\varphi}}{\ell_{\varphi}};
\ee
For $0<\kappa <2$, $\Delta\rho_{ab}$ is positive and $\Delta\rho_{c}$ negative. In other parts of the trajectories both magnetoresistivities are either positive or negative. Since the temperature dependence of the resistivities is determined by the $T-$dependence of $\ell_{\varphi}$, there is a relationship between the temperature coefficient of the resistivity and the sign of magnetoresistivity:
\bea
\frac{\partial\rho_{ab,c}}{\partial H}=Q \frac{\partial\rho_{ab,c}}{\partial
T},
\eea
where $Q=(\partial\ell_{\varphi}/\partial H )/(\partial\ell_{\varphi}/\partial T)>0$. As long as the magnetic field reduces the coherence length, its effect is equivalent to the {\it increase} in temperature. Thus, the coexistence of metallic $\rho_{ab}$ and nonmetallic $\rho_{c}$ in many cuprates translates into opposite signs of magnetoresistivities. The correlation between  MR and the sign and magnitude of $\partial\rho_{}/\partial T$ has been well documented in literature\cite{Ando,Ando1,Yan,Cimp,Cimp1}. 

There is an elegant and conceptually straightforward way to verify experimentally the relationship between the anisotropy and coherence length, Eq. (9), using the well established theory of magnetoresistance due to quantum interference. Since the coherent trajectories are confined to a single bilayer
(and therefore two-dimensional) the  field dependence of  magnetoresistance is given by
\be
\Delta\rho_{ab,c}\propto  \frac{\partial\rho_{ab,c}}{\partial T}
\ln (B/B_{\varphi})
\ee
This follows from Eqs.(42) and (43) and 
$\Delta\ell_{\varphi}\sim -\ln(B/B_{\varphi})$\cite{Lee,Cimp}. Here $B>B_{\varphi}= \hbar c/4eD\tau_{\varphi}=\phi_0/4\pi D\tau_{\varphi}$ 
and $\phi_0=\pi c\hbar/ e\approx 2\times 10^{-7}G\;cm^2$ is the quantum of flux. The coherence length in the diffusive regime is defined as $\ell_{\varphi}^2=  \langle x^2 \rangle =(1/2)\langle {\bf r}^2 \rangle = 2D\tau_{\varphi}$.  This follows from the probability to find a particle at a distance {\bf r} from the starting point: $dW({\bf r}, t )\propto \exp \{ -{\bf r}^2/4Dt\}d{\bf r}$. Thus, $B_{\varphi}= \phi_0/2\pi\ell_{\varphi}^2$.
\begin{figure*}
\includegraphics{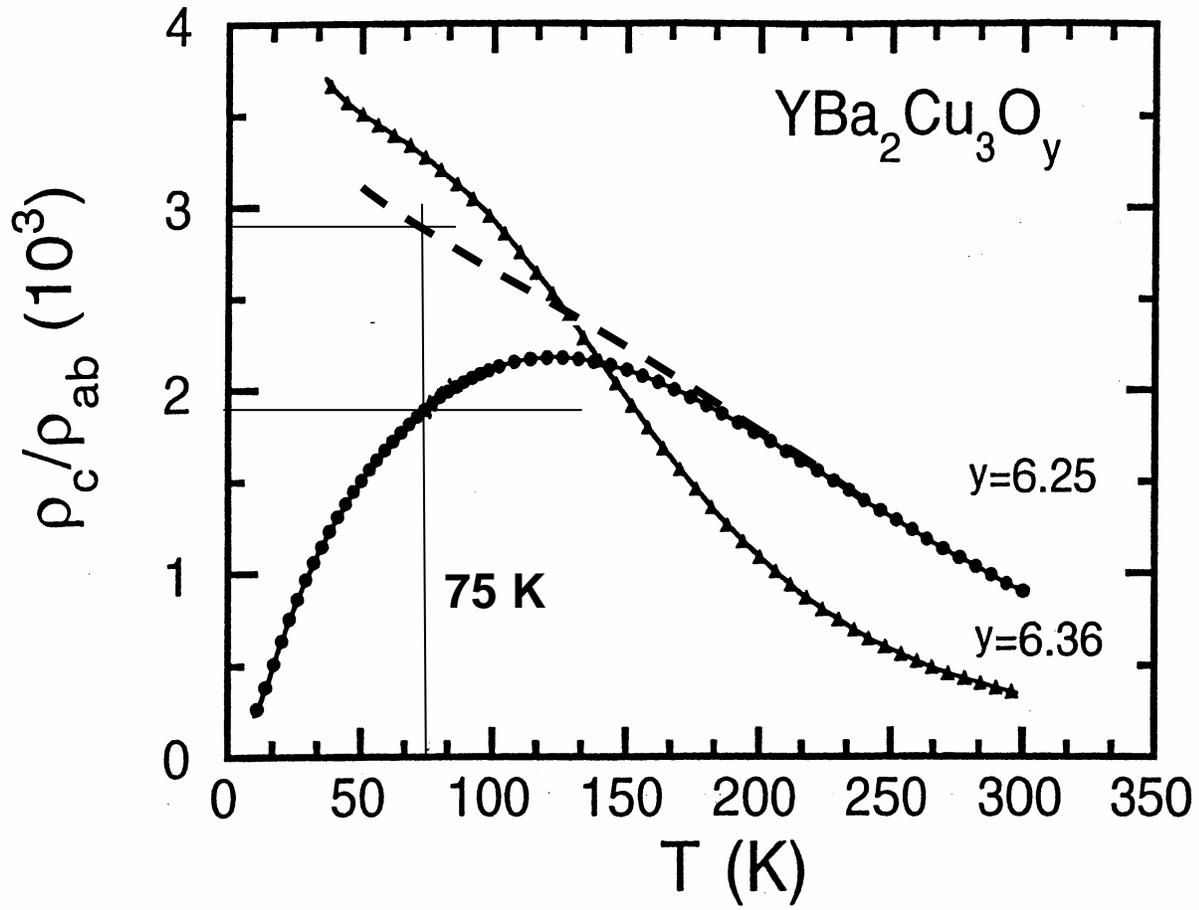}
\caption{\label{fig:wide} Temperature dependence of the anisotropy of two underdoped $YBa_2Cu_3O_{x}$ single crystals. The anisotropy of $YBa_2Cu_3O_{6.36}$ crystal increases monotonically in the temperature range shown. The more underdoped $YBa_2Cu_3O_{6.25}$ sample has non-monotonic temperature dependence of the anisotropy. The value of $\rho_c/\rho_{ab}\approx 1.9\times 10^3$ at $T=75\;K$ is indicated. The dashed curve is a hypothetical extrapolation of the high temperature behavior of the anisotropy. It shows what the values of the anisotropy would have  been, had the out-of-plane coherence length remained constant, equal to $6\AA$, see discussion in Sec. {\bf VII B}.}
\end{figure*}
According to Eq. (9), in incoherent crystals a product of seemingly totally unrelated quantities $B_{\varphi}\rho_c/\rho_{ab}$
should remain constant, even though  both the anisotropy  of resistivity and $B_{\varphi}$ strongly change with temperature. Moreover, a combination  such as
\be
\ell_{\varphi,c}=
\left (\frac{\phi_0}{2\pi B_{\varphi}}\frac{\sigma_c}{\sigma_{ab}}\right )^{1/2}
\ee
determines the out-of-plane phase coherence length and should yield
a value of the order of  interatomic (interlayer) distance. The logarithmic field dependence of MR has been observed in the normal state of
layered cuprates\cite{Cimp,Jing,Harus}. At sufficiently low temperatures
the value of $B_{\varphi}$ can be reliably established by fitting MR data to the theory of weak localization. This type of experiment requires both the anisotropy and MR to be measured on the same crystal. 

In Ref.\cite{Cimp} the magnetoresistivity data obtained on $YBa_2Cu_3O_{6.25}$ crystal at $T=75\;K$ indicate that $B_{\varphi}\approx 0.3\;T$. The temperature dependence of the anisotropy of this sample is shown in Fig. 4 and  at $T=75\;K$ it is about $1900$. Therefore, Eq.(45) yields 
\be
\ell_{\varphi,c}= 7.4\AA .
\ee
Note that in $YBa_2Cu_3O_{6.25}$ crystal the anisotropy {\it decreses} below $\approx 130\;K$. This underdoped crystal exhibits the properties of the semicoherent system, discussed below in Sec. {\bf VII}, and at $75\;K$ $\ell_{\varphi,c}$ has already increased beyond the minimum value $\ell_{0}=6\AA$. 

\subsection{\label{sec:level2} Violation of Kohler's rule}

The origin of the Kohler's rule can be illustrated on Eq.(42). In the weak field regime 
\be
\frac{\Delta\ell_{\varphi}}{\ell_{\varphi}}\sim -(\omega_c\tau_{\varphi })^2,
\ee
where $\omega_c $ is the cyclotron frequency. Thus, according to Eq.(42):
\bem
\frac{\Delta\rho_{ab}}{\rho_{ab}}\sim \kappa(\omega_c\tau_{\varphi})^2.
\eem
In Fermi liquids $\kappa\equiv 1$ and a plot of MR vs. $B^2\tau_{\varphi }^2$ produces a straight line with $T-$independent slope\cite{Luo}. In incoherent crystals even this modified Kohler's relationship does not hold because the trajectory $\kappa$ is not a constant, Fig. 1. Even if we would somehow manage to reliably determine $\tau_{\varphi}$, the slope of the plot $\Delta\rho_{ab}/\rho_{ab}$ vs. $(\omega_c\tau_{\varphi })^2$ will be proportional to $\kappa$ and change  with temperature.  

An important consequence of Eqs. (41) and (42) is that the trajectory $\kappa (T)$ or $\kappa (\sigma)$ can be extracted from MR data if the magnetoanisotropy is measured along with the components of MR. Namely, from Eqs. (41) and (42) follows: 
\be
\kappa=-\frac{2\Delta\ln \rho_{ab}}{\Delta\ln (\rho_c/\rho_{ab})}
\ee
Thus, the trajectories of the type shown in Fig. 1 or Fig. 3 can be directly obtained by measuring both components of MR similar to how it was done in Ref.\cite{Cimp}. The applicability of Eq. (48) does not depend on whether the field dependence of MR is quadratic or not.

\section{Effect of elemental substitutions and radiation-induced disorder}

The analysis given in the previous section can be readily extended to another property of layered crystals: the response of the resistivities to elemental substitutions, especially those that replace $Cu$ in the $CuO_2$ planes. 
The impurities reduce the in-plane coherence length  ($\partial\ell_{\varphi}/\partial x <0$) by decreasing the elastic mean free path and the diffusion coefficient ($x$- is the concentration of substitutions). 
Therefore, the anisotropy of the incoherent crystals always decreases with concentration of such impurities, while the resistivities change according to the position of a given crystal on the trajectory, Fig.1(a). 
\be
\frac{\partial\ln\rho_{ab}}{\partial x}=-\kappa \frac{\partial\ln
\ell_{\varphi}}{\partial x};\;\;
\frac{\partial\ln\rho_c}{\partial x}= -(\kappa -2)\frac{\partial\ln
\ell_{\varphi}}{\partial x}.
\ee
This conclusion can be illustrated on the  examples of $Bi_2Sr_2Ca(Cu_{1-x}M_x)_2O_{8+y}$ ($M=Zn,\;Mn,\;Fe,\;Co,\;$ and $Ni$)\cite{ Jeon,Sun}. Undoped $Bi_2Sr_2CaCu_2O_{8+y}$ crystals  exhibit metallic $\rho_{ab}(T)$ and nonmetallic $\rho_c(T)$, so that  below room temperature $0<\kappa <2$ and, as the result,  $\partial\rho_{ab}/\partial x >0$ while $\partial\rho_c/\partial x <0$. All elemental substitutions examined in\cite{Jeon,Sun} lead to {\it decreasing} of $\rho_c$ and {\it increasing} of $\rho_{ab}$.

Sometimes, the increase of the c-axis conductivity in response to a perturbation
has been interpreted in  literature as a crossover to coherent transport in the
c-direction. We see that this is not necessarily the case. The anisotropy and $\rho_c$ may decrease due to the reduction of the in-plane phase coherence length, even when the c-axis coherence length remains fixed.

Radiation-induced disorder in the $CuO_2$ planes may have the same effect on the resistivity as the elemental substitutions. The metallic in-plane resistivity of nonirradiated crystals and films increases with the amount of absorbed radiation and even changes its temperature dependence\cite{Tolpygo} revealing a minimum in $\rho_{ab}$ at low temperature. Unfortunately, I am not aware of any report on the effect of irradiation on the out-of-plane resistivity.  

Unlike magnetoresistivity, the response to substitutions for $Cu$ on the planes is very dramatic. Small, of the order of one percent, doping drastically affects the magnitude and temperature dependence of the resistivities. Therefore, the differential relation given by Eq.(49) can only give a qualitative idea of the changes introduced by this type of doping. For example, in $Bi_2Sr_2Ca(Cu_{1-x}Zn_x)_2O_{8+y}$ crystals just $1\%$ of $Zn$ doping reduces the anisotropy by an order of magnitude. The change in concentration of such dopants may not only drive the crystal along the given scaling trajectory, but can cause the metal-insulator transition. 

One can ask why a very small concentration of defects in the $CuO_2$ planes so drastically alters the resistivity and the in-plane coherence length, while the response to other substitutions, for example to $Pr$ substitution for $Y$, is much more gradual. One scenario is that the defects on the planes violate the reflection symmetry of the $CuO_2$ bilayers and cause hybridization of the even and odd subbands\cite{Q,L,Feng,Kordyuk}. If the relaxation rates of these subbands are very different, this hybridization may lead to drastic reduction of the decoherence time.  

\section{Semicoherent crystals}
Between two extremes, conventional Fermi liquids with temperature independent anisotropy on one hand and incoherent layered conductors, exhibiting the strongest possible rate of change of the anisotropy on the other, there is an intermediate class of layered crystals. In these systems the out-of-plane coherence length changes with temperature, but not at the same rate as the in-plane coherence length. As the result, in such ``semicoherent crystals``, the anisotropy changes with temperature, but not as strongly, and not necessarily monotonically as in incoherent layered systems. 

The first simplest scenario of semicoherence is that  $\ell_{\varphi,ab}\gg \ell_{\varphi,c}$ and
$\ell_{\varphi,c}$ is so short that the Ioffe-Regel limit still affects its temperature dependence: $\ell_{\varphi,c}=\ell_0 +v_{F,c}\tau_{\varphi}$, while $\ell_{\varphi,ab}=v_{F,ab}\tau_{\varphi}$. This empirical interpolation form of $\ell_{\varphi,c}$ takes into account that the coherence length cannot be arbitrarily small and saturates at high temperature at a finite limit.  The anisotropy then is given by
\begin{figure*}
\includegraphics{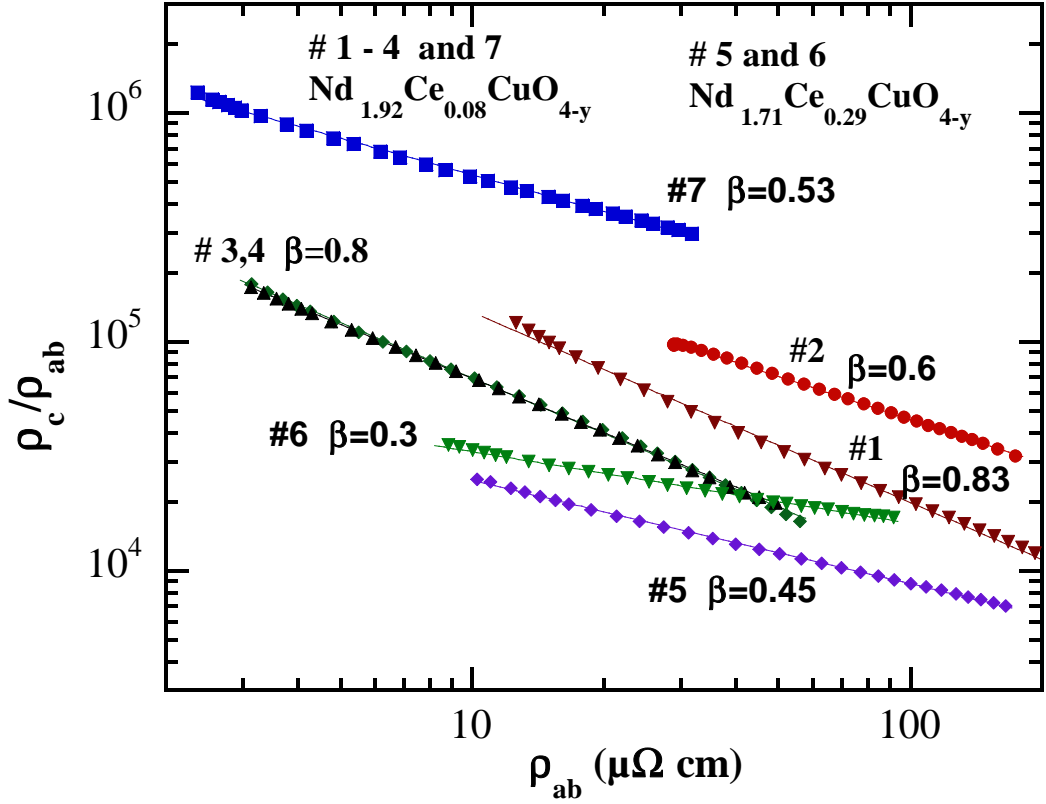}
\caption{\label{fig:wide} Anisotropy vs. in-plane resistivity for several $Nd_{2-x}Ce_xCuO_{4-y}$ crystals. The exponent $\beta$ is defined by Eq.(52). The straight lines are power law fits to the data.}
\end{figure*}

\be 
\frac{\sigma_{ab}}{\sigma_c} = \frac{v_{F,ab}^2\tau_{\varphi}^2}{(\ell_0 +v_{F,c}\tau_{\varphi})^2}.
\ee
The anisotropy monotonically increases with decreasing temperature and saturates at the level determined by the anisotropy of the three-dimensional Fermi surface. If we take the conventional form of metallic $\rho_{ab}\sim \tau_{\varphi}^{-1}$, the out-of-plane resistivity is given by:
\be
\rho_{c}=\eta^2_{max}\frac{\rho_{ab}}{(\rho_{ab}/\bar\rho +1)^2}.
\ee
It is nonmetallic at high temperatures, reaches maximum when $\rho_{ab}=\bar\rho$  and decreases at lower temperatures. There are several types of layered crystals that exhibit the temperature dependence of the resistivities qualitatively similar to that described by Eq. (51). In Ref.\cite{Valla} the coherence-incoherence transition was examined with angle-resolved photoemission spectroscopy and electronic transport measurements. In $(Bi_{0.5}Pb_{0.5})_2Ba_3Co_2O_y$ and $NaCo_2O_4$ crystals  the crossover to 3D coherence is accompanied my a maximum in $\rho_{c}(T)$, while $\rho_{ab}(T)$ remains metallic and anisotropy increases with decreasing temperature and saturates at a large value. A similar behavior of $\rho_{c}(T)$ and $\rho_{ab}(T)$ characterizes $Sr_2RuO_4$ crystals\cite{Hussey}. Note that the maximum in $\rho_{c}$ in semicoherent crystals is in stark contrast to the minimum in $\rho_{c}$ in incoherent crystals, Eqs.(20),(21) or (34). One can suggest that in some crystals it might be possible to observe the out-of-plane resistivity with two extrema: the minimum  at higher temperatures where the crystal is incoherent and the maximum at lower temperature that indicates a crossover to 3D coherence.   
\subsection{\label{sec:level2} $Nd_{2-x}Ce_xCuO_{4-y}$ crystals}

Another type of semicoherent systems that presents a greater challenge to interpretation is electron-doped  $Nd_{2-x}Ce_xCuO_{4-y}$ single crystals. These layered crystals have large anisotropy $\sim 10^4 - 10^5$ even though both resistivities $\rho_{ab}$  and $\rho_{c}$ have metallic $T-$dependence above $T_c$. The temperature dependence of the in-plane resistivity is quadratic as in conventional Fermi liquid type metals. In Ref.\cite{A2} the resitivity of several crystals with nominal $Ce$ concentration of $x=0.08$ and $0.29$ were presented. The lower and higher $Ce$ concentration corresponds to underdoped and overdoped regimes respectively. However, because of uncontrolled oxygen content the resistivities vary even within each group with the same $Ce$ content. In Fig. 5 the data from Ref.\cite{A2} are presented in order to identify the relationship between anisotropy and in-plane resistivity. For all samples it can be well approximated by:
\be
\frac{\rho_c}{\rho_{ab}}\propto \rho_{ab}^{-\beta };\;\; 
\rho_c =A\rho_{ab}^{1-\beta}.
\ee
The 3D coherent transport corresponds to the exponent $\beta =0$, while the incoherent transport is characterized by $\beta =2$ (Graf, Rainer, Sauls relationship\cite{Graf}, Eq. (11)). If we take $\rho_{ab}\sim \tau_{\varphi}^{-1}\sim\ell_{\varphi,ab}^{-1}$, Eqs. (6) and (52) are equivalent to the following relationship between the in-plane and out-of-plane coherence lengths:
\be 
\ell_{\varphi,c} \propto  \ell_{\varphi,ab}^{1-\beta/2}.
\ee
Note that in Fig. 5 the exponent $\beta$ systematically decreases with overdoping (higher $Ce$ concentration).

\subsection{Nonmonotonic temperature dependence of anisotropy}

The third scenario of gradually developing coherence in the $c-$direction differs from the previous two in the way the anisotropy changes with temperature. Both Eqs. (50) and (52) describe  monotonically increasing with decreasing temperature anisotropy and metallic in-plane resistivity. In literature one can find several examples of {\it underdoped} crystals in which $\rho_{ab}$ is non-metallic at low temperatures and anisotropy changes nonmonotonically, namely, exhibits a {\it maximum} at finite temperature. 

One example of such behavior is shown in Fig. 4. The less underdoped $YBa_2Cu_3O_{6.36}$ crystal has monotonically increasing anisotropy. The anisotropy of stronger underdoped  $YBa_2Cu_3O_{6.25}$ crystal reaches maximum at $T_m\approx 130\;K$ and decreases rapidly at lower temperatures. The temperature dependence of the resistivities of these samples is shown in Ref.\cite{Cimp}. The $YBa_2Cu_3O_{6.36}$ crystal has metallic $\rho_{ab}$,
while $YBa_2Cu_3O_{6.25}$ has nonmetallic $\rho_{ab}$ at $T<T_m$. 

According to Eq.(6), the decrease of the anisotropy at lower temperatures indicates that below $T_m$ the out-of-plane coherence length increases more rapidly than the in-plane coherence length. Presumably, if the crystal remains in the normal state at sufficiently low temperature, or superconductivity is suppressed by the magnetic field, the anisotropy of crystals like $YBa_2Cu_3O_{6.25}$ will eventually  stabilize at the level determined by the anisotropy of the 3D wave functions of the localized electrons, which is smaller than the maximum value of the anisotropy at intermediate temperatures.

There are several points worth noting. There is a widely held opinion, expressed in literature, that the tendency to establish the out-of-plane coherence increases only with increasing number density of charge carriers, while the underdoped crystals are incoherent and become more so with decreasing number of charge carriers. Especially in $YBa_2Cu_3O_{y}$, the disruption  of the $CuO$ chains is thought to be the main reason for increasing anisotropy. This is obviously not the case, and the systematic study of the evolution of the anisotropy with oxygen content\cite{ Euro} shows that the anisotropy at a given temperature reaches maximum at an intermediate oxygen content. 

Apparently, in cuprates there is a mechanism of interlayer decoherence, of still unknown nature, which can be overcome by the phase transition into the superconducting state. This takes place in optimally and slightly underdoped cuprates. If, however, the onset of superconductivity is somehow prevented, one can expect that at sufficiently low temperatures 3D coherence will still be established. As in the case of $YBa_2Cu_3O_{6.25}$, Fig. 4, this will manifest itself as the decrease in anisotropy below a certain temperature.  How low is this  temperature is a measure of strength of the decoherence mechanism. Then, it is obvious that the mechanism of decoherence is {\it stronger} in $YBa_2Cu_3O_{6.36}$ and other $YBCO$ crystals with higher oxygen content than in $YBa_2Cu_3O_{6.25}$. This observation has to be contrasted with the behavior of $Pr$ doped, {\it fully oxygenated}, $Y_{1-x}Pr_xBa_2Cu_3O_{7-\delta}$ single crystals\cite{Levin,Euro}. Even the very strongly underdoped $Y_{0.47}Pr_{0.53}Ba_2Cu_3O_{7-\delta}$ and insulating $PrBa_2Cu_3O_{7-\delta}$ remain incoherent, with increasing anisotropy, at temperature as low as $1.9\;K$. 

   A conclusion that one can make from this is that in $YBCO$ and its derivatives the $CuO$ chain layer is responsible for the interlayer decoherence. In $Y_{1-x}Pr_xBa_2Cu_3O_{7-\delta}$ crystals  the chain layer is fully formed and the mechanism of decoherence is strongest. In $YBa_2Cu_3O_{6.25}$ this mechanism is weakened to such extend that it allows the process of establishing 3D coherence to begin at temperature as high as $130\;K$. 

The central role of the chain layer in $YBCO$ is also corroborated by the value of $\ell_{\varphi,c}\approx 7.4\;\AA$, Eq.(46). At this temperature ($75\;K$) the anisotropy $\rho_c/\rho_{ab}\approx 1900$ and  $\ell_{\varphi,c}$ is already increasing beyond its minimum value $\ell_{0}\approx 6\AA$ at higher  temperatures. 
If $\ell_{\varphi,c}$ were still equal $6\AA$ at $T=75\;K$ the anisotropy at this temperature would have been $1900\times(7.4/6)^2=2900$. This is close to the value of the anisotropy one gets extrapolating the high temperature behavior of the anisotropy of $YBa_2Cu_3O_{6.25}$ as shown by the dashed line in Fig. 4. 

Thus, the loss of phase coherence of the charge carriers takes place during their transfer from the $CuO_2$ planes to the charge reservoir. Otherwise, if the loss of coherence was taking place during the charge carriers transfer between the neighboring $CuO_2$ bilayers, the value of $\ell_{0}$ would be equal to the distance between the bilayers ($12\AA$) (see Section III and comparison with the solvable model\cite{Graf}).
\subsection{\label{sec:level2} Intrinsic tunneling spectroscopy in semicoherent crystals}
The intrinsic tunneling spectroscopy in small crystals\cite{Kleiner,Kleiner1} and mesa-structures\cite{ Schlenga,Suzuki,Krasnov,Yurgens,Yurgens2} may provide a powerful tool for investigation of the process of establishing the interlayer coherence of single charge carriers. The most studied crystals such as $Bi_2Sr_2CaCu_2O_{8+x}$ ($Bi-2212$) and $Bi_2Sr_2Cu_2O_{6+\delta}$ ($Bi-2201$) are found to have current-voltage (I-V) characteristics similar to that of a stack of Josephson junctions. These intrinsic junctions are formed by the charge reservoir layers sandwiched between superconducting $CuO_2$ units (a superconducting unit is either a single $CuO_2$ layer like in $Bi-2201$ or a $CuO_2$ bi-layer in $Bi-2212$ and $YBCO$). Below critical temperature the I-V characteristics consist of multiple hysteretic branches. Each branch corresponds to switching of one intrinsic junction from the superconducting to the normal state. 

In mesa structures the number of charge reservoir layers can be made small ($10-20$) and, correspondingly, the number of branches is easily countable.  For example, in Ref.\cite{Krasnov} the number of branches remains the same in the temperature interval from $70$ to $4.2\;K$ ($Bi-2212$). This is understandable if the single quasiparticles remain incoherent, so that the Cooper pairs from each individual $CuO_2$ unit form a Josephson junction across every charge reservoir in the mesa. The normal state $c-$axis resistance of that sample increases with decreasing temperature, corroborating the assertion that this crystal is incoherent. 

What would change in this picture if the crystal was semicoherent? If the out-of-plane phase coherence length of single quasiparticles $\ell_{\varphi,c}$ is already greater than its minimum value $\ell_0$ at temperatures above $T_c$, but at $T_c$ is still shorter than the height of the mesa, the number and the properties of the intrinsic junctions will be different from those formed in incoherent crystal. To illustrate that, let us say that just above $T_c$, in the normal state, a number of neighboring $CuO_2$ units have formed "coherent clusters". This means that the single charge quasiparticles in these clusters consist of the orbitals of several (two or more) neighboring $CuO_2$ units, as opposed to incoherent crystals where the wavefunctions of the quasiparticles consist of the orbitals of one $CuO_2$ unit only. The rest of the $CuO_2$ units in the mesa remain incoherent. 

Let $N$ be the number of $CuO_2$ units in the mesa. And let $n_2$ be the temperature dependent number of double coherent clusters ($n_2<N/2$). Then $N-2n_2$ is the number of single incoherent units. For simplicity we have neglected the probability of formation of coherent clusters consisting of three and more $CuO_2$ units. 

Below $T_c$ the quasiparticles form Cooper pairs in both coherent clusters and single units. When the superconducting order parameter is extended throughout the crystal, the intrinsic junctions with the lowest critical current $J_c$ are the ones between two neighboring coherent clusters, or between a single unit and coherent cluster, or between two single units. The total number of these junctions is $N-n_2(T)$. In other words, a charge reservoir "hidden" inside every double cluster does not form a weak junction. Thus, the number of hysteretic branches in I-V characteristic $N_b=N-n_2(T)$ and, therefore, can be noticeably smaller than the number of unit cells along the c-axis.  Even more telling, the number of branches will be temperature dependent, decreasing with decreasing temperature. 

Generally, we can introduce $n_k$ as  the average number of coherent clusters consisting of $k$ $CuO_2$ units so that $\sum_{k=1}^{N}kn_k=N$. The number of charge reservoir layers hidden inside these clusters is $\sum_{k=1}^{N}(k-1)n_k$. Then the number of low$-J_c$ hysteretic junctions and, correspondingly, the number of low$-J_c$ hysteretic branches $N_b= \sum_{k=1}^{N}n_k$ - the total number of coherent clusters. In incoherent crystals $N_b=N$, since the only clusters are single units. As the coherence of single charge quasiparticles sets in, the number of coherent clusters along the height of the mesa decreases and in the limit when $\ell_{\varphi,c}$ exceeds the height of the mesa, the mesa itself becomes one coherent cluster so that only one branch of $I-V$ characteristic, $V=0$, remains as long as $I<J_c$. 

Obviously, this a very simplistic description of a complex process of breaking up of coherence along the long chain of $CuO_2$ units. For example, one can expect that once the lowest$-J_c$ junctions are broken, the increasing current will start breaking coherence of the coherent clusters. This may result in another sequence of hysteretic branches. Joule heating will complicate further this picture. My point, however, is that in the mesas made of semicoherent crystals the pattern of hysteretic branches and its evolution with temperature might be distinctly different from that found in $Bi-2201$ or $Bi-2212$. 

Semicoherent crystals for this type of experiment, besides those already mentioned above, may include $RBa_2Cu_3O_{6+x}$ ($R=Tm,Lu$)\cite{Lavrov} and $Nd_{2-x}Ce_xCuO_{4-y}$\cite{Onose}. A characteristic feature that indicates semicoherence is the maximum in $\rho_c$ or anisotropy at $T=100-150\;K$.

\section{conclusions and suggestions}

This paper presents a phenomenological approach to understanding the normal state transport properties of incoherent and semicoherent layered crystals. It is based on the fundamental relationship between the resistive anisotropy and the phase coherence lengths derived in Sec. II. This relationship is especially useful when one of the coherence lengths is fixed, temperature independent distance as is the case in some layered crystals such as superconducting cuprates and many others. In Sec. III we have shown that the results obtained by our approach are equivalent to those obtained for a solvable microscopic model\cite{Graf}. Application of this approach to hopping conduction in layered crystals and to crystals like optimally doped cuprates show a good agreement with experimental data. 

In incoherent crystals the resistive anisotropy is a measure of the in-plane coherence length and, therefore, such systems allow an effective application of the scaling theory as demonstrated in Sec. IV. One of the important consequences of the scaling approach is the idea of the universal trajectories. We suggest that the families of cuprates such as $YBa_2Cu_3O_{6+y}$ with different oxygen content or $Y_{1-x}Pr_xBa_2Cu_3O_{7-\delta}$ with different $Pr$ concentration are described by the same dependence of conductivity vs. anisotropy, Eqs. (23),(24). This type of scaling has been found in Ref\cite{Euro}.

Moreover, the scaling approach allows to predict two important asymptotic branches of the in-plane conductivity. Using simple analytical form of the scaling trajectory we have shown that the metallic branch of incoherent crystal can be described by logarithmically increasing {\it conductivity}: $\sigma_{ab}\propto \ln (\ell_{\varphi})$. This asymptotic behavior is illustrated in Fig. 2. For the insulating branch, using a simple analytical expansion of the scaling trajectory, we obtain the {\it resistivity} that is increasing logarithmically: $\rho_{ab}\propto \ln (\ell_{\varphi})$. At sufficiently low temperatures, when  $\ell_{\varphi}\sim T^{-\gamma}$, this crosses over into the famous logarithmic temperature dependence: $\rho_{ab}\propto \ln (1/T)$.

In Sec. V the magnetoresistivity of incoherent crystals is discussed. One of the most interesting conclusions is that at temperatures where MR due to quantum interference can be observed one can also determine the out-of-plane phase coherence length, Eq.(45).  In Sec. VI we consider the effect of elemental substitutions and radiation induced disorder. The treatment used in the previous sections also allows to explain sometimes puzzling conflicting responses of in-plane and out-of-plane resistivities to the perturbations such as impurities and imperfections. 

In Sec. VII we discussed the semicoherent crystals in which both in- and out-of-plane coherence lengths change with temperature, but at different rates. Three scenarios are discussed; all of them, it seems, can be found in either overdoped or underdoped cuprates. 

\subsection{Possible future experiments}

Typically in literature the problem of the normal state of the cuprates is stated in terms of the anomalies of the out-of-plane conductivity. We see that the properties of $\sigma_c$ are entirely determined by the fact that the out-of-plane coherence length has a fixed value and does not change with temperature, Eq. (10). Then, the real unresolved question about the anomalous properties of the normal state in cuprates is the nature of the strong mechanism of decoherence of the charge carriers over very short distance (confinement). While this question is important in and of itself, it is also a key question which is necessary to address in order to understand the superconductivity in cuprates, which is likely to be the phase transition that allows to overcome the confinement. 

One possible scenario of decoherence was outlined in Ref.\cite{Turlakov}, where the confinement was attributed to the dephasing of the tunneling charge carriers resulted from the interaction with charge fluctuations. While this effect appears to be too weak to account for the phenomenon of confinement, it raises a question: how strong is the dephasing mechanism? For example, as was discussed above, $Pr$ doped $YBCO$ crystals remain incoherent even at $T\approx 2\;K$. However, if the confinement is the result of a conventional process of finite strength, one can reasonably assume that eventually, at sufficiently low temperature, even  these systems will begin to form 3D coherence in the normal state. 

As was discussed in Sec. {\bf VII B} the decoherence mechanism appears to be  determined by the properties of the charge reservoir and is weaker in both overdoped and underdoped cuprates. In underdoped cuprates the transition to 3D coherence manifests itself as the maximum of anisotropy at finite temperature which has been observed in YBCO, Fig. 4, and in $Nd_{2-x}Ce_xCuO_{4-y}$ crystals\cite{Onose}. However, the strongest confinement in $YBCO$ is associated with fully oxygenated $CuO$ chain layer. In my view, it would be very important to find out if, indeed, the fully oxygenated derivatives of $YBCO$, in which superconductivity is suppressed, will exhibit the maximum in anisotropy in the normal state. This can be done with $Pr$ doped $YBCO$ or $YBa_2(Cu_{1-x}M_x)_3O_{7-\delta}$, or a combination of both types of doping optimized with the goal to minimize $T_c$, but retain reasonably high conductivity. 

Another important experiment would be to verify the Eq.(45). Since the theory of MR due to quantum interference is well established, this relationship is a direct consequence of our main result, Eqs. (6) and (9). At this point it has been applied at only one temperature and the result looks reasonable, Eq. (46). If it can be substantiated for different types of incoherent crystals that the length determined by Eq. (45) is indeed of the order of interlayer distance and remains relatively $T-$independent over a wide range of temperatures, it will be an experimental proof of the fundamental relationship between the resistive anisotropy and coherence lengths.  
 
There remains also the question of the universal trajectories, Eq. (23). In Ref.\cite{Euro} the two parameter scaling dependences $\sigma(\eta)$, Eq. (24), were  obtained by measuring the resistivities $\rho_c$ and $\rho_{ab}$. However, the magnetoresistivity measurements allow to go further and obtain the one parameter differential trajectories $\kappa (\sigma)$ directly according to Eq. (48). 

Last but not least is the possibility to observe the establishing of interlayer coherence with the help of intrinsic tunneling spectroscopy. To my knowledge, no experiments on the crystals in which $\rho_c(T)$ exhibits a crossover from incoherent to coherent transport have been conducted so far. 
\\
                                                                              
Acknowledgment: Part of this research was performed while the author held a National Research Council Senior Research Associateship Award at the Air Force Research Laboratory.


\begin{references}
\bibitem{Cooper} S.L. Cooper and K.E. Gray, in Physical Properties
of High Temperature Superconductors IV, edited by D. M. Ginsberg
(World Scientific, Singapore, 1994).
\bibitem{Anderson} P. W. Anderson, The Theory of Superconductivity in the
High-$T_c$ Cuprates, Princeton University Press, (1997).
\bibitem{H} J. H. Kim et al.  Physica (Amsterdam){\bf247C},297(1995)


\bibitem{Tsvetkov} A. A. Tsvetkov et al. Nature {\bf 395}, 360 (1998)
\bibitem{Basov} D. N. Basov et al. Science {\bf 283}, 49 (1999).
\bibitem{Marel} D. van der Marel et al. Physica (Amsterdam){\bf 341C-248C}, 1531 (2000).
\bibitem{Qiu} X. G. Qiu et al. Physica (Amsterdam){\bf 341C-248C}, 1383 (2000).
\bibitem{Graf} M. J. Graf, D. Rainer, and J. A. Sauls, Phys. Rev. B {\bf 47}, 12089 (1993).
\bibitem{Abrahams} E. Abrahams, P.W. Anderson, D.C. Licciardello,
and T.V. Ramakrishnan, Phys. Rev. Lett. {\bf 42}, 673 (1979).
\bibitem{Thouless} D.J. Thouless, Phys. Rep. 13C, 93 (1974).
\bibitem{Gant} V. F. Gantmakher et al. JETP Lett. 65, 870 (1997)
\bibitem{Ono} S. Ono et al. Phys. Rev. Lett. {\bf 85}, 638 (2000).
\bibitem{Lee} P. A. Lee and T.V. Ramakrishnan, Rev. Mod. Phys. 57, 287 (1985).
\bibitem{Economu} E. N. Economu, Green's Functions in Quantum Physics 2nd Edition, Springer 1983 (p. 153). 
\bibitem{Th} D. J. Thouless, in Ill-Condensed Matter, Edited by R. Balian, R. Maynard, and G. Toulouse, North-Holland (1979). 
\bibitem{Akkermans} E. Akkermans and G. Montambaux, Phys. Rev. Lett. {\bf 68}, 642 (1998).
\bibitem{Abrikosov} A. A. Abrikosov, Fundamentals of the Theory of Metals, North-Holland (1988). 
\bibitem{Levin} G. A. Levin, T. Stein, C.C. Almasan, S. H. Han, D. A. Gajewski, and M. B. Maple, Phys. Rev. Lett. {\bf 80}, 841 (1998).
\bibitem{Shklovskii} B. I. Shklovskii and A. L. Efros, Electronic properties of Doped Semiconductors, Springer-Ferlag (1984) p.152.
\bibitem{A1} C.C. Almasan, E. Cimpoiasu, G. A. Levin, H. Zheng, A. P. Paulikas, and B. W. Veal,  J. Low. Temp. Phys. 117, 1307 (1999); cond-mat/9908233.
\bibitem{Friedmann} T.A. Friedmann et al. Phys. Rev. B {\bf 42}, 6217 (1990).
\bibitem{Ando} Y. Ando, G. S. Boebinger, A. Passner, T. Kimura, and K. Kishio,
Phys. Rev. Lett. {\bf 75}, 4662 (1995).
\bibitem{Ando1} Y. Ando, G. S. Boebinger, A. Passner, N. L. Wang, C.
Geibel, and F. Steglich, Phys. Rev. Lett. {\bf 77}, 2065 (1996).
\bibitem{Boebinger} G. S. Boebinger et al. Phys. Rev. Lett. {\bf 77}, 5417 
(1996).
\bibitem{Fournier} P. Fournier et al. Phys. Rev. Lett. {\bf 81}, 4720 (1998).
\bibitem{Krasnov} V. M. Krasnov, A. Yurgens, D. Winkler, P. Delsing, and T. Claeson, Phys. Rev. Lett. 84, 5860 (2000)
\bibitem{Yurgens} A. Yurgens, D. Winkler, T. Claeson, S. Ono, and Y. Ando, Phys. Rev. Lett. 90, 147005 (2003)


\bibitem{Yurgens1} A. Yurgens, D. Winkler, T. Claeson, S. Ono, and Y. Ando, cond-mat/0309131.
\bibitem{Yang} G. Yang, J. S. Abell, and C. E. Gough, Appl. Phys. Lett. {\bf 75}, 1955 (1999).
\bibitem{Takenaka} K. Takenaka et al. Phys. Rev. B 50, 6534 (1994).
\bibitem{Euro} G. A. Levin, E. Cimpoiasu, H. Zheng,  A. P. Paulikas, B. W. Veal, Shi Li,  M. B. Maple, and  C. C. Almasan, Europhysics Letters 57, 86 (2002). 
\bibitem{Kravchenko}  S. V. Kravchenko and T. M. Klapwijk, Phys. Rev. Lett. 84, 2909 (2000).
\bibitem{Dobro} V. Dobrosavljevic et al. Phys. Rev. Lett. 79, 455 (1997).
\bibitem{Cimp} E. Cimpoiasu, G. A. Levin, C. C. Almasan, Hon Zheng, and B. W. Veal,  Phys. Rev. B 63, 104515 (2001).
\bibitem{Harris} J. M. Harris et al. Phys. Rev. Lett. {\bf 75}, 1391 (1995).
\bibitem{Cimp1} E. Cimpoiasu, G. A. Levin, C. C. Almasan, A.P. Paulikas, and B. W. Veal, Phys. Rev. B 64, 104514 (2001).
\bibitem{JAP} G. A. Levin J. Appl. Phys. 81, 714 (1997).
\bibitem{Yan} Y. F. Yan, P. Matl, J. M. Harris, and N. P. Ong, Phys. Rev. B
{\bf 52}, R751 (1995).
\bibitem{Jing} T. W. Jing et al. Phys. Rev. Lett. 67, 761 (1991).
\bibitem{Harus} G. I. Harus et al. JETP Letters 70, 97 (1999).
\bibitem{Luo} Nie Luo and G. H. Miley, Physica (Amsterdam) 371C, 259 (2002)
\bibitem{Sun} X. F. Sun et al. Phys. Rev. B 62, 11384 (2000).
\bibitem{Jeon} D-S. Jeon, M. Akamatsu, H. Ikeda, and R. Yoshizaki, Physica (Amsterdam) 253C, 102 (1995).
\bibitem{Tolpygo} S. K. Tolpygo et al.  Phys. Rev. B{\bf 53}, 12462 (1996).

\bibitem{Q} K. F. Quader and G. A. Levin,  Phil. Mag. B 74, 611 (1996).
\bibitem{L} G. A. Levin and  K. F. Quader, Phys. Rev. B{\bf 62}, 11879 (2000).
\bibitem{Feng} D. L. Feng et al. Phys. Rev. Lett. 86, 5550 (2002) 
\bibitem{Kordyuk} A. A. Kordyuk et al. Phys. Rev. Lett. 89, 077003 (2002).
\bibitem{Valla} T. Valla et al. Nature 417, 627 (2002).
\bibitem{Hussey} N. E. Hussey et al.  Phys. Rev. B{\bf 57}, 5505 (1998).
\bibitem{A2}C. C. Almasan, G. A. Levin, E. Cimpoiasu, T. Stein, C. L. Zhang, M. C. de Andrade, M. B. Maple, H. Zheng, A. P. Paulikas, and B. W. Veal,  Int. J. Mod. Phys. B 13, 3618 (1999); cond-mat/9910016. 
\bibitem{Kleiner} R. Kleiner, F. Steinmeyer, G. Kunkel, and P. Muller, Phys. Rev. Lett. 68, 2394 (1992).
\bibitem{Kleiner1}  R. Kleiner and P. Muller, Phys. Rev. B 49, 1327 (1994).
\bibitem{Schlenga} K. Schlenga et al. Phys. Rev. B 57, 14518 (1998).
\bibitem{Suzuki} M. Suzuki, T. Watanabe, and A. Matsuda, Phys. Rev. Lett. 82, 5361 (1999).
\bibitem{Yurgens2} A. A. Yurgens, Supercond. Sci. Technol. 13, R85 (2000).  
\bibitem{Lavrov} A. N. Lavrov, M. Yu. Kameneva, and L. P. Kozeeva, Phys. Rev. Lett. {\bf 81}, 5636 (1998). 
\bibitem{Onose} Y. Onose, Y. Taguchi,  K. Ishizaka and Y. Tokura,  Phys. Rev. B 69, 024504 (2004). 
\bibitem{Turlakov} M. Turlakov and A. J. Leggett, Phys. Rev. B 63, 064518 (2001). 
\end{references}

\end{document}